\documentclass[12pt, theoremstly, screen, oneside, en]{minimalthesis}

\usepackage[all, -amssymb]{mathlib}
\usepackage{graphicx}
\usepackage{verbatim}
\usepackage{enumitem}
\numberwithin{equation}{section}

\usepackage{bbm}
\usepackage{pdfpages}

\newcommand{\la}{\ensuremath\langle}
\newcommand{\ra}{\ensuremath\rangle}
\newcommand{\eq}[1]{\begin{align}#1\end{align}}
\newcommand{\Tr}{\ensuremath\mathrm{Tr}}
\newcommand{\Lemma}[1]{\begin{lem}#1\end{lem}}
\newcommand{\Def}[1]{\begin{defn}#1\end{defn}}
\newcommand{\Theorem}[2][]{\begin{thm}[#1] #2\end{thm}}
\newcommand{\Bem}[1]{\begin{rem}#1\end{rem}}
\newcommand{\HS}{\ensuremath\mathcal{L}^2(\mathfrak{h})}
\newcommand{\err}{\ensuremath \mathcal{O}\big(\|\delta z\|_{\HS}^2\big)}
\newcommand{\bounded}{\ensuremath\mathfrak{B}(\mathfrak{h})}
\newcommand{\Real}{\ensuremath\mathrm{Re}}
\newcommand{\Bog}{\ensuremath\mathrm{Bog}_J[\mathfrak{h}]}
\usepackage{hyperref}
\usepackage{csquotes, xpatch}
\usepackage[backend = bibtex, style=numeric]{biblatex}
\usepackage{nomencl}

\begin{document}

\begin{titlepage}
	\begin{center}
	\vspace*{1cm} 
	      
    \Huge
    \textbf{Minimization of an Energy Functional for an Electron in the quantized Radiation
Field over quasifree States}
	\vspace*{4cm}
	
	\large
	 handed in on April 24, 2023 by 
	\vspace*{0.5cm}
	\\ \Large Matthias Herdzik
	\vspace*{3cm}
	
	\vfill
	\Large
	Technische Universität Braunschweig
	\vspace*{0.15cm}
	
	\large
	Master of Science in Mathematics\\
	Supervisor: Prof. Dr. Volker Bach\\
	Second assessor: Prof. Dr. Dirk Lorenz\\
	\end{center}
\end{titlepage}
\pagenumbering{Roman}

\newpage

\includepdf{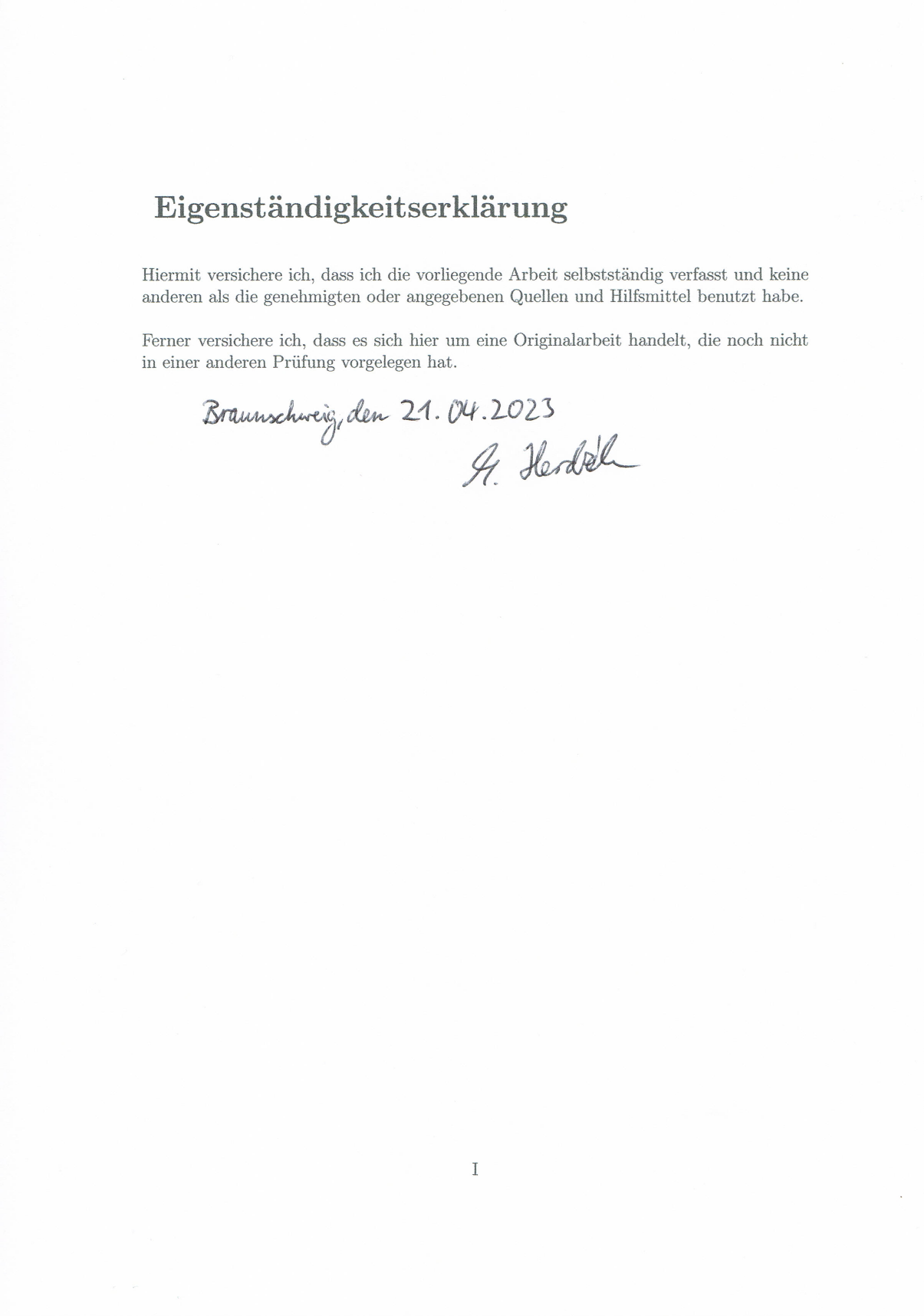}

\newpage
\tableofcontents
\newpage
\pagenumbering{arabic}

\section{Introduction}

In \cite{BBT} the authors studied the Bogolubov-Hartee-Fock energy of the Hamiltonian 
\eq{
H_{g,\vec{p}}=\frac{1}{2} \big(\vec{P}_f+\vec{\mathbb{A}}(\vec{0})-\vec{p}\big)^2 +H_f,
} meaning its infimum 
on the set of quasifree density matrizes. It is shown [3, Theorem~IV.5] that the Bogolubov-Hartree Fock energy coincides with the infimum over \textit{pure} quasifree states and that there is a unique minimizier within this class if $|\vec{p}|$ is sufficiently small. 
Additionally, a necessary condition on the minimizer is given in the form of Euler-Lagrange equations [3, Theorem VIII.8]. The Hamiltonian $H_{g,\vec{p}}$ can be viewed as a pointwise decompositon in $\vec{p}$ of the Pauli-Fierz Hamiltonian 
\eq{\tilde{H_g}:=\frac{1}{2}\big(\tfrac{1}{i}\vec{\nabla}-\vec{\mathbb{A}}(\vec{x})\big)+H_f,
}
which describes a non-relativistic, spinless particle coupled to the quantized radiation field and is the primary object of study. Note that information on the spectrum of $\tilde{H}_g$ can be retrieved from knowledge about the spectra of each $H_{g,\vec{p}}$ [6, Theorem~XIII.85]. The Pauli-Fierz Hamiltonian is a self-adjoint  operator defined on a subset of the tensor product $L^2(\mathbb{R}^3) \otimes \mathfrak{F}_b(\mathfrak{h})$, with $L^2(\mathbb{R}^3)$ being the Hilbert space of electron wave functions and $\mathfrak{F}_b(\mathfrak{h})$ being the Hilbert space of wave functions of an unknown number of photons, the so called photon Fock space. The Lieb-Loss approximation, which is studied in \cite{BH} deals with the minimization of the Pauli-Fierz Hamiltonian on the set of density matrices that can be written as pure tensor products $\phi_{el}\otimes \psi_{ph}$, where $\phi_{el} \in L^2(\mathbb{R}^3)$ and $\psi_{ph}\in \mathfrak{F}_b(\mathfrak{h})$ are normed. Of particular interest for us is the derivation of [4, Theorem IV.5], which requires various techniques that we will be applying as well. 
\\
\\The mathematical framework is presented in Section II. In Section III the expectation value of $H_{g,\vec{p}}$ in a pure quasifree state is calculated as a functional of the Bogolubov transformation that defines it. The same expectation value has already been calculated in \cite{BBT} but as a functional of generalized one-particle density matrices instead. In Section IV we use a technique similar to the prove of (\cite{BH}, Lemma IV.8) in order to relate the Bogolubov-Hartree-Fock energy to minimization over pure quasifree density matrices with certain additional symmetries.
In Section V we slightly modify the parametrization used in the proof of (\cite{BH}, Lemma IV.9), so that the resulting functional can be easily differentiated in Section VI. A discussion of the results can be found in Section VII.


\newpage
\section{Prerequisites}

The following mathematical framework is taken from a variety of sources and at the end of each section information on the literature may be found.
We assume knowledge that may be acquired in an introductory course on functional analysis as given. 
\\
\\Throughout the entirety of this text Hilbert spaces are always assumed to be complex, separable Hilbert spaces.

\subsection{Trace Class and Hilbert-Schmidt Operators}

In order to define density matrices we first need to understand the notion of trace in infinite dimensional Hilbert spaces. The usual trace of a matrix $A\in \mathbb{R}^{N\times N}$ for some $N\in\mathbb{N}$ is just the sum of its diagonal elements \eq{\Tr(A):=\sum_{n=1}^N \la \varphi_n | A \varphi_n\ra,
}
where $\{\varphi_n\}_{n=1}^N\subseteq \mathbb{R}^N$ is an orthonormal basis. 
In infinite dimensional Hilbert spaces however, we need to think about the convergence of (II.1) as well. First, we define the trace for positive operators. We recall that an operator $A\in \bounded$ is \textbf{positive} if for any $\varphi\in\mathfrak{h}$ we have $\la \varphi|A \varphi \ra\geq 0.$

\Def{
Let $\mathfrak{h}$ be a Hilbert space and let $\{\varphi_n\}_{n=1}^\infty\subseteq \mathfrak{h}$ be an ONB. For a positive operator $A\in\bounded$ the \textbf{trace of $A$} is defined as \eq{
\Tr(A):=\sum_{n=1}^\infty \la \varphi_n | A \varphi_n\ra.
}
}
\noindent Note that the trace is independent of the choice of the orthonormal basis $\{\varphi_n\}_{n=1}^\infty\subseteq \mathfrak{h}$. In order to define the trace for a more general class of operators we need the absolute value of a bounded operator.

\Def{
Let $A\in\bounded$. Then we define the \textbf{absolute value of $A$} by \eq{|A|:=\sqrt{A^\star A}.}
}

\Bem{
The absolute value of operators lacks many of the properties that we are accustomed to from the absolute value of complex numbers. For example the identities $|AB|=|BA|$, $|AB|=|A||B|$ and $|A^\star|=|A|$ do not hold true in general. However, for positive operators it is true that $|A|=A$.
}

\Def{
An operator $A\in \bounded$ is called \textbf{trace class} if and only if
$
\Tr(\:|A|\:)< \infty.
$ The family of trace class operators is denoted by $\mathcal{L}^1(\mathfrak{h}).$
}

\Bem{
The trace class operators $\mathcal{L}^1(\mathfrak{h})$ become a Banach space when equipped with the norm $\|\cdot\|_{\mathcal{L}^1}:=\Tr[\:|\cdot|\:]$. Furthermore $\mathcal{L}^1(\mathfrak{h})$ is a two-sided ideal of $\bounded$ and the adjoint of a trace class operator remains trace class.
}

\noindent For every trace class operator the trace can now be defined as in (II.2).

\Theorem{
Let $A\in \mathcal{L}^1(\mathfrak{h})$ and let $\{\varphi_n\}_{n=1}^\infty\subseteq\mathfrak{h}$ be an orthonormal basis. Then $\sum_{n=1}^\infty \la \varphi_n | A \varphi_n \ra $ converges absolutely and the limit is independent of the choice of basis.
}

\noindent The most important properties of the trace are summarized in the next theorem.

\Theorem{
The map $\Tr:\mathcal{L}^1(\mathfrak{h})\to \mathbb{C}, A\mapsto \Tr(A):=\sum_{n=1}^\infty \la \varphi_n | A \varphi_n \ra $, where $\{\varphi_n\}_{n=1}^\infty\subseteq\mathfrak{h}$ is an orthonormal basis, fulfills the following properties.
\begin{enumerate}[label=(\roman*)]
\item $\Tr(\cdot)$ is linear.
\item $\Tr (A^\star) = \overline{\Tr (A)}$ if $A\in \mathcal{L}^1(\mathfrak{h})$.
\item $\Tr (AB)=\Tr (BA)$ if $A\in \mathcal{L}^1(\mathfrak{h})$ and $B\in \bounded$.
\end{enumerate}
}

A similar class of operators is the class of Hilbert-Schmidt operators. 

\Def{
An operator $A\in\bounded$ is called \textbf{Hilbert-Schmidt} if $\Tr[A^\star A] < \infty$. The family of all HIlbert-Schmidt operators is denoted by $\HS$.
}

\Bem{
The Hilbert-Schmidt operators become a Hilbert space when equipped with the inner product $\la A|B\ra_{\HS}:= \Tr (A^\star B)$. The Hilbert-Schmidt operators $\HS$ are also a two-sided ideal in $\bounded$.
}

\noindent We collect a few inequalities that are useful for dealing with trace class and Hilbert-Schmidt operators in the next theorem.
\newpage
\Theorem{
The following inequalities hold true.
\begin{enumerate}[label=(\roman*)]
\item Let $A,B\in\HS$. Then the Cauchy-Schwarz inequality for Hilbert-Schmidt operators assumes the form \eq{
|\Tr( A^\star B)| \leq \Tr (A^\star A)^\frac{1}{2} \:\Tr(B^\star B)^\frac{1}{2}=\|A\|_{\HS}\|B\|_{\HS}.}
\item If $A\in \mathcal{L}^1(\mathfrak{h})\cap \HS$, then \eq{\|A\|_{op}\leq \|A\|_{\HS} \leq\|A\|_{\mathcal{L}^1(\mathfrak{h})}.}
\item Let $A\in \mathcal{L}^1(\mathfrak{h})$ and $B\in \bounded$. Then $BA\in\mathcal{L}^1(\mathfrak{h})$ and \eq{
|\Tr(BA)|\leq \|BA\|_{\mathcal{L}^1(\mathfrak{h})} \leq \|B\|_{op} \|A\|_{\mathcal{L}^1(\mathfrak{h})}.}
\end{enumerate}
}

\noindent Finally, we can define the set of density matrices.

\Def{
A trace class operator $\rho\in \mathcal{L}^1(\mathfrak{h})$ is called a \textbf{density matrix} if and only if $\rho=\rho^\star \geq 0$ and $\Tr(\rho)=1$. We denote the set of density matrices by $\mathfrak{DM}$. If $\rho$ has rank one, then it is called a \textbf{pure density matrix}.
}

\Bem{
Every vector in $\mathfrak{h}$ defines a pure quasifree density matrix through its projection.
Let $\rho = |\psi\ra\la\psi|$ with $\psi\in\mathfrak{h},\|\psi\|=1$ and let $\{\varphi_n\}_{n=1}^\infty \subseteq \mathfrak{h}$ be an orthonormal basis with $\varphi_1:=\psi$. Then we obtain that
\eq{
\Tr(\rho A) = \sum\nolimits_{n=1}^\infty \la \varphi_n | \psi \ra \la \psi | A \varphi_n\ra 
=\la \psi| A \psi \ra,
} assuming for example $A\in \bounded$.
}

\underline{References:} The majority of this section is taken from Chapter VI.6 of \cite{RSI}. The definition of density matrices and the following remark can similarly be found in \cite{HFT}.

\subsection{Unbounded Operators}

In comparison to bounded operators, the study of unbounded operators poses a lot more technical difficulties. By the Hellinger-Toeplitz theorem [5, p.84] for example we know that a symmetric, unbounded operator cannot be defined on all of $\mathfrak{h}$.

\Def{
Let $\mathfrak{h}$ be a Hilbert space. Let $\mathcal{D}\subseteq \mathfrak{h}$ be a linear subspace and let $A:\mathcal{D}\to \mathfrak{h}$ be a linear map. The pair $(A,\mathcal{D})$ is called a \textbf{linear operator on} $\mathfrak{h}$ and in this case $\mathcal{D}$ is called the \textbf{domain of $A$}. The set of linear operators is denoted by $\mathfrak{L}(\mathfrak{h})$. 
}

\noindent We introduce the following notions.

\Def{
Let $\big(A,\mathcal{D}(A)\big)\in \mathfrak{L}(\mathfrak{h})$ be a linear operator.
\begin{enumerate}[label=(\roman*)]
\item $\big(A,\mathcal{D}(A)\big)$ is called \textbf{densely defined} if $\mathcal{D}(A)$ is a dense set in $\big(\mathfrak{h},\la\cdot|\cdot\ra\big)$.
\item The linear subspace 
\eq{
\mathcal{G}\big(A,\mathcal{D}(A)\big):=\big\{(x,Ax)\:|\: x\in \mathcal{D}(A)\big\}\subseteq \mathfrak{h}\oplus \mathfrak{h}
} is called the \textbf{graph of $\big(A,\mathcal{D}(A)\big)$}.
\item A linear operator $\big(B,\mathcal{D}(B)\big)\in \mathfrak{L}(\mathfrak{h})$ is called \textbf{extension of $\big(A,\mathcal{D}(A)\big)$} if \eq{
\mathcal{G}\big(A,\mathcal{D}(A)\big) \subseteq \mathcal{G}\big(B,\mathcal{D}(B)\big).
}
\item $\big(A,\mathcal{D}(A)\big)$ is called \textbf{closed} if $\mathcal{G}\big(A,\mathcal{D}(A)\big)\subseteq \mathfrak{h}\oplus \mathfrak{h}$ is a closed subset.
\item $\big(A,\mathcal{D}(A)\big)$ is called \textbf{closeable} if the closure of $\mathcal{G}\big(A,\mathcal{D}(A)\big)$ is the graph of an operator $\big(\overline{A},\mathcal{D}\big(\overline{A}\big)\big)\in \mathfrak{L}(\mathfrak{h})$. In this case $\big(\overline{A},\mathcal{D}\big(\overline{A}\big)\big)$ is called the \textbf{closure of $A$}. 
\end{enumerate}
}

\noindent A closed unbounded operator fulfills a property similar to continuity despite the fact that it is not continuous at any point. Given a closed operator $\big(A,\mathcal{D}(A)\big)$ and let $x,y,x_1, x_2,...\in\mathfrak{h}$ such that $x_n\to x$ and $Ax_n\to y$ as $n\to\infty$. Then $y$ must be in $\mathcal{D}(A)$ and $Ax=y$.
Also note that being densely defined is necessary for the definition of the adjoint through the Riesz lemma.

\Def{
Let $\big(A,\mathcal{D}(A)\big)\in \mathfrak{L}(\mathfrak{h})$ be a densely defined operator. 
\begin{enumerate}[label=(\roman*)]
\item Let $D(A^\star)$ be the set of $x\in\mathfrak{h}$ for which there is a $z\in \mathfrak{h}$ such that
\eq{
\forall y\in \mathcal{D}(A):\la x | Ay\ra = \la z | y\ra.
} For each such $x\in \mathcal{D}(A^\star)$ define $A^\star x=z$. The resulting operator $\big(A^\star, \mathcal{D}(A^\star)\big)$ is called the \textbf{adjoint of $\big(A,\mathcal{D}(A)\big)$}.
\item $\big(A,\mathcal{D}(A)\big)$ is called \textbf{symmetric} if \eq{
\forall x,y\in\mathcal{D}(A): \la x|Ay\ra = \la Ax|y\ra.
}
\item $\big(A,\mathcal{D}(A)\big)$ is called \textbf{self-adjoint} if \eq{
\big(A,\mathcal{D}(A)\big)=\big(A^\star,\mathcal{D}(A^\star)\big).
}
\item $\big(A,\mathcal{D}(A)\big)$ is called \textbf{essentially self-adjoint} if its closure is self-adjoint.
\end{enumerate}
}
\noindent We emphasize that in (iii) the domains of $A$ and $A^\star$ have coincide $\mathcal{D}(A)=\mathcal{D}(A^\star)$.
The adjoint has the following properties.

\Theorem{
Let $\big(A,\mathcal{D}(A)\big)\in\mathfrak{L}(\mathfrak{h})$ be a densely defined operator. \begin{enumerate}[label=(\roman*)]
\item The adjoint $\big(A^\star,\mathcal{D}(A^\star)\big)$ is a closed operator.
\item $\big(A,\mathcal{D}(A)\big)$ is closable if and only if $\mathcal{D}(A^\star)$ is dense.
\item $\big(A,\mathcal{D}(A)\big)$ is symmetric if and only if $\big(A^\star,\mathcal{D}(A^\star)\big)$ is an extension of $\big(A,\mathcal{D}(A)\big)$. Therefore a symmetric operator is always closable.
\end{enumerate}
}

\noindent Self-adjoint operators can be diagonalized similarly to hermitian matrices. This property is formulated in the spectral theorem.

\Theorem[Spectral Theorem]{
Let $(A,\mathcal{D})\in\mathfrak{L}(\mathfrak{h})$ be a self-adjoint operator. Then there is a measure space $(M,\mu, \mathfrak{A})$ with $\mu$ a finite measure, a unitary operator $U:\mathfrak{h}\to L^2(M,\mathrm{d}\mu)$, and a real-valued function $f$ on $M$ which is finite a.e. so that 
\begin{enumerate}[label=(\roman*)]
\item $\psi\in\mathcal{D}$ if and only if $f(\cdot)(U\psi)(\cdot)\in L^2 (M,\mathrm{d}\mu),$
\item If $\varphi \in U[\mathcal{D}]$, then $
(UAU^\star \varphi)(m)=f(m)\varphi(m)$ for $m\in M$.
\end{enumerate}
}
\noindent The Spectral Theorem is an extremly useful tool. It also enables the definition of a functional calculus.

\Def{
Let $(A,\mathcal{D})\in\mathfrak{L}(\mathfrak{h})$ be a self-adjoint operator and let $(M,\mu, \mathfrak{A})$, $U$ and $f$ be as given by the spectral theorem. Let $g:\mathbb{R}\to \mathbb{C}$ be measureable and bounded. Then we define the operator $g(A)\in\bounded$ through\eq{
\forall \varphi \in L^2(M,\mathrm{d}\mu): (Ug(A)U^\star \varphi)(m)=g(f(m))\varphi(m).
}
}
\noindent Note that $\|g(A)\|_{op}=\sup_{\lambda\in\sigma(A)} |g(\lambda)|\leq \|g\|_{L^\infty}.$
Another useful notion in the study of unbounded operators is relative boundedness.

\Def{
Let $\big(A,\mathcal{D}(A)\big), \big(B, \mathcal{D}(B)\big)\in\mathfrak{L}(\mathfrak{h})$ be densely defined. $\big(B, \mathcal{D}(B)\big)$ is called $A$-bounded if $\mathcal{D}(B)\supseteq \mathcal{D}(A)$ and there exist $0<a,b<\infty$ such that 
\eq{
\forall \varphi \in \mathcal{D}(A): \|B\varphi\|\leq a \|A\varphi\| +b \|\varphi\|.
}
}
\Bem{
If $\big(B, \mathcal{D}(B)\big)$ is $A$-bounded, then $B\in\mathfrak{B}[\mathcal{D}(A);\mathfrak{h}]$, where $\mathcal{D}(A)$ is equipped with the graph norm $\|\varphi\|_A:=\|\varphi\|+\|A\varphi\|$. 
}

\underline{References:} The material from this section can be found in both Chapter VIII of \cite{RSI} and Chapters I and III of \cite{RT}. The above presentation is a mixture of the two.

\subsection{Fock Space}
The Fock space is introduced when the number of particles is either uncertain or changing. While its construction looks intimidating at first glance, it is just the direct sum of the $N$-particle spaces. We can obtain a probability for the system having $N$ particles by projecting a Fock space vector into the $N$-particle subspace and calculating the square of the norm.

\Def{
Given a Hilbert space $\mathfrak{h}$ we define the \textbf{Fock Space $\mathcal{F}(\mathfrak{h})$ of $\mathfrak{h}$} as the direct sum
\eq{
\mathcal{F}(\mathfrak{h}) = \bigoplus_{N=0}^\infty \mathfrak{h}^{\oplus N},
}
where $\mathfrak{h}^{\oplus N}$ is the N-fold tensor product of $\mathfrak{h}$ with itself if $N\geq 2$ and 
\eq{\mathfrak{h}^{\oplus 0}:= \mathbb{C}, \mathfrak{h}^{\oplus 1} := \mathfrak{h}.
} We write $\mathfrak{h}^{\otimes 0} := \mathrm{span}\{\Omega\}$ and call $\Omega \in \mathbb{C}, |\Omega|=1$ the vacuum vector. 
}

\Bem{
The N-particle space $\mathfrak{h}^{\otimes N}$ is treated as subspace of $\mathcal{F}(\mathfrak{h})$ by identifying $\psi_N\in\mathfrak{h}^{\otimes N}$ with the vector \eq{
(\underbrace{0,\dots,0}_{\text{N times}},\psi_N,0,\dots)\in\mathcal{F}(\mathfrak{h}).
}}

\noindent Since we intent to describe photons, we want to make use of their bosonic symmetries. The projection onto the symmetric vectors describes invariance under exchange of particles. The bosonic Fock space therefore contains only vectors that remain invariant when particles are exchanged.

\Def{
Let $\mathfrak{h}$ be a Hilbert space. The \textbf{projection onto the symmetric vectors} is the operator $P = \bigoplus_{N=0}^\infty P_N \in \mathfrak{B}[\mathcal{F}(\mathfrak{h})]$ where $P_0=\mathrm{id}_\mathbb{C}$, $P_1:=\mathrm{id}_\mathfrak{h}$ and for all $N\geq 2$ $P_N$ is the linear and continuous extension of
\eq{
\forall \varphi_1,\dots, \varphi_N \in \mathfrak
{h}: P_N(\varphi_1\otimes\cdot\cdot\cdot\otimes\varphi_N) := \frac{1}{N!} \sum_{\sigma\in\mathcal{S}_N} \varphi_{\sigma(1)}\otimes \cdot\cdot\cdot \otimes \varphi_{\sigma(N)}
}
where $\mathcal{S}_N$ is the symmetric group of $\{1,\dots,N\}$.
}

\Def{
Given a Hilbert space $\mathfrak{h}$ we define the \textbf{bosonic Fock space} as the subspace
\eq{
\mathcal{F}_b(\mathfrak{h}):=P[\mathcal{F}(\mathfrak{h})]= \bigoplus_{N=0}^\infty \mathcal{F}_b^{(N)}(\mathfrak{h}) \subseteq \mathcal{F}(\mathfrak{h}),
} where $\mathcal{F}_b^{(N)}(\mathfrak{h})=P_N[\mathfrak{h}^{\otimes N}].$
}
\Bem{
The projection onto the symmetric vectors $P$ is indeed an orthogonal projection. Therefore the bosonic Fock space $\mathcal{F}_b(\mathfrak{h})$ is a closed subspace of $\mathcal{F}(\mathfrak{h})$.
}
\Bem{
The N-boson space $\mathcal{F}_b^{(N)}(\mathfrak{h})$ is treated as subspace of $\mathcal{F}_b(\mathfrak{h})$ by identifying $\psi_N\in\mathcal{F}_b^{(N)}(\mathfrak{h})$ with the vector \eq{
(\underbrace{0,\dots,0}_{\text{N times}},\psi_N,0,\dots)\in\mathcal{F}_b(\mathfrak{h}).
}
}

\underline{References:}
This introduction on Fock spaces is heavily inspired by their presentation in Chapter 7 of \cite{Sol}, however, there are influences from Chapter VI of \cite{RT} as well.

\subsection{Bosonic Creation and Annihilation Operators}
In the next three sections we introduce operators on Fock spaces. The creation and annihilation operators can be thought of as adding or removing a particle.

\Def{
Let $\mathfrak{h}$ be a Hilbert space. We define 
\begin{enumerate}[label=(\roman*)]
\item the \textbf{number operator} $\big(\mathcal{N},\mathcal{D}(\mathcal{N})\big)\in \mathfrak{L}[\mathcal{F}(\mathfrak{h})]$ as 
\eq{
\mathcal{N}:=\bigoplus_{N=0}^\infty N\cdot \mathrm{id}_{\mathfrak{h}^{\otimes N}},
}
and its domain is defined by
\eq{
\mathcal{D}(\mathcal{N}):=\big\{\Psi=(\psi_N)_{N=0}^\infty\in \mathcal{F}(\mathfrak{h}): \sum_{N=0}^\infty N^2 \|\psi_N\|^2 < \infty\big\}.
}
\item for any $\varphi\in \mathfrak{h}$ the \textbf{creation operator} $\big(\tilde{a}^\star(\varphi),\mathcal{D}\big(\mathcal{N}^\frac{1}{2}\big) \big)\in \mathfrak{L}[\mathcal{F}(\mathfrak{h})]$ by $\tilde{a}^\star(\varphi)(\Omega)=\varphi$ and its action on tensor products \eq{
\forall \varphi_1,\dots,\varphi_N \in \mathfrak{h}: \tilde{a}^\star(\varphi)(\varphi_1\otimes \cdot\cdot\cdot \otimes \varphi_N):=(N+1)^\frac{1}{2} \varphi\otimes \varphi_1\otimes \cdot\cdot\cdot \otimes \varphi_N,
} if $N\geq 1$. The domain is defined as
\eq{
\mathcal{D}\big({\mathcal{N}^\frac{1}{2}}\big):=\Big\{\Psi=(\psi_N)_{N=0}^\infty\in \mathcal{F}(\mathfrak{h}): \sum\nolimits_{N=0}^\infty N \|\psi_N\|^2 < \infty\Big\}.
}
\item for any $\varphi\in \mathfrak{h}$ the \textbf{annihilation operator} $\big(\tilde{a}(\varphi),\mathcal{D}\big(\mathcal{N}^\frac{1}{2}\big) \big)\in \mathfrak{L}[\mathcal{F}(\mathfrak{h})]$ by $\tilde{a}(\varphi)(\Omega)=0$ and its action on tensor products \eq{
\forall \varphi_1,\dots,\varphi_N \in \mathfrak{h}: \tilde{a}(\varphi)(\varphi_1 \otimes \cdot\cdot\cdot \otimes \varphi_N):= N^\frac{1}{2} \la \varphi | \varphi_1 \ra  \varphi_2 \otimes \cdot\cdot\cdot \otimes \varphi_N,
} if $N\geq 1$. 
\end{enumerate}
}

\noindent It is worthwhile to introduce creation and annihilation operators for the bosonic case. Aside from mapping the symmetric vectors into themselves, they fulfill the canonical commutation relations presented in Theorem II.6. 

\Def{
Let $\mathfrak{h}$ be a Hilbert space. We define \begin{enumerate}[label=(\roman*)]
\item for any $\varphi \in \mathfrak{h}$ the \textbf{bosonic creation operator} $\big(a^\star(\varphi),P\mathcal{D}\big(\mathcal{N}^\frac{1}{2}\big)\big)\in \mathfrak{L}\big(\mathcal{F}_b(\mathfrak{h})\big) $ by
\eq{
a^\star(\varphi):=P\circ \tilde{a}^\star(\varphi) \circ P, and
}
\item for any $\varphi \in \mathfrak{h}$ the \textbf{bosonic annihilation operator} $\big(a(\varphi), P\mathcal{D}\big(\mathcal{N}^\frac{1}{2}\big)\big)\in \mathfrak{L}\big(\mathcal{F}_b(\mathfrak{h})\big) $ by
\eq{
a(\varphi):=P\circ \tilde{a}(\varphi) \circ P.}
\end{enumerate}
}
\Bem{
The bosonic creation operator $a^\star(\varphi)$ is linear in $\varphi$. The map $\varphi\mapsto a(\varphi)$ however is antilinear. We recall that a map $f$ between two complex vector spaces $X$ and $Y$ is called \textbf{antilinear} if \eq{
\notag (i)&\:\forall x_1,x_2\in X : f(x_1+x_2)= f(x_1)+f(x_2),
\\ (ii)&\: \forall x\in X, \forall \alpha\in\mathbb{C}: f(\alpha x ) = \overline{\alpha} f(x).
}
}

\Theorem{
Let $\mathfrak{h}$ be a Hilbert space. The bosonic creation and annihilation operators fulfill the \textbf{canonical commutation relations (CCR)}:
\eq{
\notag (i)&\: \forall \varphi_1,\varphi_2\in \mathfrak{h}: [a^\star(\varphi_1),a^\star(\varphi_2)]=[a(\varphi_1),a(\varphi_2)]=0,
\\(ii)& \: \forall \varphi_1,\varphi_2\in \mathfrak{h}: [a(\varphi_1),a^\star(\varphi_2)]=\la\varphi_1 | \varphi_2 \ra \cdot \mathrm{id}_{\mathcal{F}_b(\mathfrak{h})},
}
Furthermore they are a \textbf{Fock representation} of the CCR, which means that \eq{\forall \varphi \in \mathfrak{h}: a(\varphi)\Omega=0.}
}

At this point we present the quasifree density matrices as defined in [3, Section II.2 and Definition III.6].

\Def{
A density matrix $\rho\in\mathfrak{DM}$ is called \textbf{quasifree}, if there exist $f_\rho\in\mathfrak{h}$, a symplectomorphism $T_\rho$ and a positive, self-adjoint operator $h_\rho=h_\rho^\star \geq 0$ on $\mathfrak{h}$ such that 
\eq{
\Tr_\mathfrak{F}\:[\rho W(\sqrt{2}f/i)]= \exp \big[2i\: \mathrm{Im}\la f_\rho | f\ra - \tfrac{1}{2}\la T_\rho f | (1+h_\rho) T_\rho f \ra\big],
}
for all $f\in\mathfrak{h}$, where 
\eq{
W(f):=\exp \big[\tfrac{i}{\sqrt{2}}\big(a^\star(f)+a(f)\big)\big].
}
A symplectomorphism $T$ is a continuous, $\mathbb{R}$-linear automorphism on $\mathfrak{h}$ such that\eq{
\forall f,g\in \mathfrak{h}: \mathrm{Im}\la Tf | Tg \ra = \mathrm{Im}\la f|g\ra.
}
}

\underline{References:} Creation and annihilation operators are discussed in both Chapter 7 of \cite{Sol} and Chapter VI of \cite{RT}. The notation is influenced by both of these source, but we made modifications in order to suit the calculations that we make later on.

\subsection{Generalized Creation and Annihilation Operators}

For some applications the bosonic creation and annihilation operators are quite impractical. In Section III we will compute expectation values for strings of creation and annihilation operators, that is concatenations $a^\#(\varphi_1)\cdot\cdot\cdot a^\#(\varphi_k)$, where $a^\#(\varphi)$ is either $a^\star (\varphi)$ or $a(\varphi)$ and $\varphi_1,\dots,\varphi_k\in\mathfrak{h}, k\in\mathbb{N}.$ Instead of calculating each of these $2^k$ expectation values depending on whether each $a^\#$ is a creation or annihilation operator, we only have to calculate one expectation value when using the generalized operators.

\Def{
An antilinear bijection $J:\mathfrak{h}\to \mathfrak{h}$ is called an \textbf{antiunitary involution} if $J^2=\mathrm{id}_{\mathfrak{h}}$ and
\eq{\forall \varphi_1,\varphi_2\in \mathfrak{h} : \la J\varphi_1 | J \varphi_2 \ra = \la \varphi_2 |\varphi_1 \ra.
}
Furthermore for an antiunitary involution $J$ define the real-linear map 
\eq{
q:\mathfrak{h}\to \mathfrak{h}\oplus\mathfrak{h}, \varphi \mapsto (\varphi,J\varphi)^T.
}
}

\Bem{
For elements of $\mathfrak{h}\oplus \mathfrak{h}$ the notations $(f,g)^T$ and $f\oplus g$ with $f,g\in \mathfrak{h}$ are used interchangebly.
}
\newpage
\Def{
Let $J:\mathfrak{h}\to \mathfrak{h}$ be an antiunitary involution. For $f\oplus g \in \mathfrak{h} \oplus \mathfrak{h}$ we define 
\begin{enumerate}[label=(\roman*)]
\item the \textbf{generalized creation operator} $A^\star_J\in\mathfrak{B}\big[\mathcal{D}\big(\mathcal{N}^\frac{1}{2}\big);\mathcal{F}_b(\mathfrak{h})\big]$ by \eq{A^\star_J(f\oplus g):= a^\star(\varphi)+a(Jg),} 
\item the \textbf{generalized annihilation operator} $A_J\in\mathfrak{B}\big[\mathcal{D}\big(\mathcal{N}^\frac{1}{2}\big);\mathcal{F}_b(\mathfrak{h})\big]$ by \eq{A_J(f\oplus g):= a(\varphi)+a^\star(Jg).}
\end{enumerate}
}

\Bem{
The generalized creation and annihilation operators are also refered to as \textbf{field operators}.
}

\noindent In the following theorem two important properties of the field operators are presented. The first property is precisely the reason why we do not need to distinguish between creation and annihilation operators when calculating expectation values of strings of field operators $A^\#_J(F_1)\cdot \cdot\cdot A^\#_J(F_k)$. Instead we can calculate only $A_J(F_1)\cdot \cdot\cdot A_J(F_k)$ and use (II.38) to rewrite the appearing generalized creation operators as generalized annihilation operators. The second property resembles the CCR however in general $[A(F_1),A(F_2)]$ and $[A^\star(F_1),A^\star(F_2)]$ do not vanish. 

\Theorem{
Let $J:\mathfrak{h}\to \mathfrak{h}$ be an antiunitary involution. Then the following statements hold true.
\begin{enumerate}[label=(\roman*)]
\item Let $\mathcal{J}:={\big(\begin{smallmatrix}
0 & J \\
J & 0
\end{smallmatrix}\big)}\in \mathfrak{B}(\mathfrak{h}\oplus \mathfrak{h}) $. Then \eq{\forall F \in \mathfrak{h}\oplus \mathfrak{h} : A_J(F) = A^\star_J(\mathcal{J}F).}
\item Let $\mathcal{S}:={\Big(\begin{smallmatrix}
\mathrm{id}_\mathfrak{h} & 0\\
0 & -\mathrm{id}_\mathfrak{h}
\end{smallmatrix}\Big)}\in \mathfrak{B}(\mathfrak{h}\oplus \mathfrak{h})$. Then
\eq{
\forall F_1,F_2\in\mathfrak{h}\oplus\mathfrak{h}: [A_J(F_1),A_J^\star(F_2)]=\la F_1 | \mathcal{S} F_2 \ra.
}
\end{enumerate}
}

\underline{References:} This presentation of the field operators is taken from Chapter III of \cite{BH}.

\subsection{Second Quantization}
The second quantization is a way of obtaining an operator on Fock space from a one-particle  operator. For example, the dynamics of a free particle is given by the Schrödinger equation through the Hamiltonian $\big(-\Delta, H^2(\mathbb{R}^3)\big)$. If $N\geq2$ free particles are involved, the N-particle Hamiltonian assumes the form $\sum_{n=1}^N \mathrm{id}_{\mathfrak{h}^{\otimes n-1}} \otimes (-\Delta_{x_n}) \otimes \mathrm{id}_{\mathfrak{h}^{\otimes N-n}}$ and acts on the N-particle space $H^2(\mathbb{R}^3)^{\otimes N}$. We would like for the Hamiltonian on Fock space to act like the N-particle Hamiltonian on the N-particle subspace for every $N\in\mathbb{N}$.

\Def{
Let $(\omega, \mathcal{D}) \in \mathfrak{L}[\mathfrak{h}]$ be essentially self adjoint. Then the \textbf{second quantization $\mathrm{d}\Gamma(\omega)$ of $\omega$} is the operator on $\mathcal{F}(\mathfrak{h})$ defined by 
\eq{
\mathrm{d}\Gamma(\omega):=0_\mathbb{C}\oplus \omega\oplus \bigoplus_{N=2}^\infty \sum_{j=1}^N \omega_j^{(N)},
} where $\omega_j^{(N)}:= \mathrm{id}_{\mathfrak{h}^{\otimes j-1}} \otimes \omega \otimes \mathrm{id}_{\mathfrak{h}^{\otimes N-j}}$.
}

\noindent A useful technical tool is the subspace of finite vectors since its elements are a lot easier to deal with than arbitrary Fock space vectors. Oftentimes it suffices to show a statement on the subspace of finite vectors and argue that it holds true in general. It is also a suitable domain for the second quantization.

\Def{
Let $\mathfrak{h}$ be a Hilbert space and let $\mathcal{D}\subseteq \mathfrak{h}$ be a dense subspace. Let \eq{\mathfrak{F}_{fin}^{(N)}(\mathcal{D}):=\mathrm{span}\big\{\varphi_1\otimes\cdot\cdot\cdot \otimes \varphi_N \big|\varphi_1,\dots,\varphi_N\in \mathcal{D}\big\}.
} The \textbf{subspace of finite vectors over $\mathcal{D}$} is defined by
\eq{
\notag \mathfrak{F}_{fin}(\mathcal{D}):=\bigcup_{N=1}^\infty \Big\{\big(&\phi^{(0)},\phi^{(1)},\dots,\phi^{(N)},0,0\dots\big)\Big| \\ &\phi^{(0)}\in \mathfrak{F}_{fin}^{(0)}(\mathcal{D}),
\phi^{(1)}\in \mathfrak{F}_{fin}^{(1)}(\mathcal{D}),\dots,\phi^{(N)}\in \mathfrak{F}_{fin}^{(N)}(\mathcal{D})\Big\}.
}
}

\Theorem{
Let $(\omega,\mathcal{D}) \in \mathfrak{L}[\mathfrak{h}]$ be essentially self-adjoint. Then the second quantization $ \big(\mathrm{d}\Gamma(\omega),P[\mathfrak{F}_{fin}(\mathcal{D})]\big)\in \mathfrak{L}[\mathfrak{F}_b(\mathfrak{h})]$ is essentially self-adjoint and obeys the following identity for any ONB $\{\varphi_k\}_{k=1}^\infty\subseteq\mathcal{D}$ of $\mathfrak{h}$
\eq{
\mathrm{d}\Gamma(\omega)=\sum_{k,l=1}^\infty \la \varphi_k |\omega \varphi_l\ra a^\star(\varphi_k)a(\varphi_l).
}
}

\Bem{
The authors of \cite{RT}, \cite{BBT} and \cite{BH} employ an integral representation of the second quantization \eq{
\mathrm{d}\Gamma(\omega)=\int \omega(k) a^\star(k) a(k) \mathrm{d}k,
} that will not be used throughout this text. 
}

\noindent Inspired by (II.43) we can define the second quantization of a bounded operator on $\mathfrak{h}\oplus\mathfrak{h}$  by using generalized creation and annihilation operators.

\Def{
Let $T\in\mathfrak{B}(\mathfrak{h}\oplus \mathfrak{h})$ be self-adjoint, $y\in \mathfrak{h}$ and $J:\mathfrak{h}\to \mathfrak{h}$ an antiunitary involution. The \textbf{second quantization $\mathrm{d}\Gamma_J[T,y]\in\mathfrak{B}[\mathcal{D}(\mathcal{N});\mathfrak{F}_b(\mathfrak{h})]$ of $T$} is defined as
\eq{
\notag \mathrm{d}\Gamma_J[T,y]:=&\sum\nolimits_{i,j=1}^\infty \la F_i|TF_j\ra A^\star_J(F_i)A_J(F_j) \\&+ \sum\nolimits_{i=1}^\infty \big\{\la F_i|q(y)\ra A^\star_J(F_i) + \overline{\la F_i| q(y)\ra} A_J^\star(F_i)\big\},
} where $\{F_i\}_{i=1}^\infty\subseteq \mathfrak{h}\oplus \mathfrak{h}$ is any orthonormal basis. 
}

\Theorem{
Let $T\in\mathfrak{B}(\mathfrak{h}\oplus \mathfrak{h})$ be self-adjoint, $y\in \mathfrak{h}$ and $J:\mathfrak{h}\to \mathfrak{h}$ an antiunitary involution. Then the second quantization $\big(\mathrm{d}\Gamma_J[T,y], \mathcal{D}(\mathcal{N})\big) \in\mathfrak{L}(\mathfrak{h})$ is self-adjoint.
}

\underline{References:}
For more information on the introductory example, see Chapter 2.1 of \cite{Sol}. Definition II.19 is inspired by Chapter VI of \cite{HFT}. The presentation of the subspace of finite vectors and Theorem II.8 is taken from a remark in Chapter VI of \cite{RT}. The second quantization in terms of field operators can be found in Chapter III.2 of \cite{BH}.

\subsection{Bogolubov Transformations}

Another advantage of generalized field operators is that it makes defining and working with Bogolubov transformations much easier. We define Bogolubov transformations through their action on field operators. Other equivalent characterizations may be found in [3, Lemma IV.2].

\Def{
Let $\mathfrak{h}$ be a Hilbert space and let  $J:\mathfrak{h}\to\mathfrak{h}$ be an antiunitary involution. 
\begin{enumerate}[label=(\roman*)]
\item The \textbf{set of generators of Bogolubov transformations} is defined as
\eq{
\mathrm{Bog}_J[\mathfrak{h}]:=\Big\{B=\begin{pmatrix}
U & JVJ\\
V & JUJ
\end{pmatrix} \Big| B^\star \mathcal{S}B=\mathcal{S}, \Tr(V^\star V)<\infty \Big\},
} where $\mathcal{S}={\big(\begin{smallmatrix}
1 & 0\\
0 & -1
\end{smallmatrix}\big)}$.
\item A \textbf{homogeneous Bogolubov transformation} is a unitary operator $\mathbb{U}_B \in \mathfrak{B}[\mathfrak{F}_b(\mathfrak{h})]$ that acts on the field operators as
\eq{
\forall F\in \mathfrak{h}\oplus \mathfrak{h} : \mathbb{U}_B A(F) \mathbb{U}_B^\star = A(BF),
}where $B\in\Bog$.
\item A \textbf{Weyl transformation} is a unitary operator $\mathbb{W}_\eta \in  \mathfrak{B}[\mathfrak{F}_b(\mathfrak{h})]$ that acts on the field operators as
\eq{
\forall F \in \mathfrak{h}\oplus \mathfrak{h}: \mathbb{W}_\eta A(F) \mathbb{W}_\eta^\star =A(F)+ \overline{\la q(\eta) | F\ra},
} where $\eta \in \mathfrak{h}$. 
\item A \textbf{Bogolubov transformation} is a composition of homogeneous Bogolubov transformations and Weyl transformations.
\end{enumerate}
}

\Bem{The requirement $\Tr(V^\star V)<\infty$ is calles Shale-Stinespring condition.
The requirement $B^\star \mathcal{S} B=\mathcal{S}$ implies the following relations between $U,V$ and~$J$
\eq{
\notag U^\star U -V^\star V &= \mathrm{id},\\
\notag U^\star JV &= V^\star JU,\\
UU^\star -JVV^\star J &= \mathrm{id},\\
\notag UJV^\star &= VU^\star J.
}
}
\noindent The second quantization $\mathrm{d}\Gamma_J$ has simple transformation properties under Bogolubov transformations.

\Theorem{
Let $B\in\Bog$ and $\eta\in \mathfrak{h}$. The second quantization $\mathrm{d}\Gamma_J[T,y]$ transforms under Bogolubov transformations according to 
\eq{
\mathbb{U}_B \mathrm{d}\Gamma_J[T,y] \mathbb{U}_B^\star = \mathrm{d}\Gamma_J[BTB^\star,\tfrac{1}{2}q^\star B q(y)],
}
and
\eq{
\mathbb{W}_\eta \mathrm{d}\Gamma_J[T,y]\mathbb{W}_\eta^\star = \mathrm{d}\Gamma_J[T,y+\tfrac{1}{2} q^\star T q(\eta)] +\la \eta | q^\star T q(\eta)\ra +4\Real \la \eta |y\ra.
}
}

Particularly helpful for the computation of expectation values of strings of field operators in a quasifree state is the Wick Theorem.

\Theorem[Wick Theorem]{
Let $\Psi=\mathbb{U}_B^\star \Omega$ and $F_1,\dots,F_{2m}\in \mathfrak{h}\oplus \mathfrak{h}$, for $m\geq 1$. Then \eq{
\notag \la \Psi | &A_J(F_1)\cdot\cdot\cdot A_J(F_{2m})\Psi\ra =\\& \sum_{\sigma\in P_{2m}} \la \Psi | A_J(F_{\sigma(1)})A_J(F_{\sigma(2)})\Psi\ra \cdot\cdot\cdot \la \Psi | A_J(F_{\sigma(2m-1)})A_J(F_{\sigma(2m)}) \Psi\ra,
}
where \eq{\notag P_{2m}:=\{\sigma\in\mathcal{S}_{2m}| \sigma(2j-1&)<\sigma(2j+1),j=1,\dots,m-1, \\&\sigma(2j-1)<\sigma(2j),j=1,\dots,m \}} is the set of pairings and \eq{
\la\Psi|A_J(F_1)\cdot\cdot\cdot A_J(F_{2m-1}) \Psi\ra =0.
}
}

\underline{References:} The above presentation on Bogolubov transformations can be found in Chapter III.2 of \cite{BH}. The Wick Theorem is taken from Chapter 10 of \cite{Sol}. 

\subsection{The Pauli-Fierz Hamiltonian}

The Pauli-Fierz Hamiltonian is the linear operator
\eq{
\tilde{H}_{g} := \frac{1}{2}\big(i\vec{\nabla}+\vec{\mathbb{A}}(\vec{x})\big)^2+H_f \label{Pauli-Fierz Hamiltonian}
} defined on its domain $H^2(\mathbb{R}^3)\otimes \mathcal{D}(\mathcal{N})\subseteq L^2(\mathbb{R}^3)\otimes \mathfrak{F}_b(\mathfrak{h})$. The one-photon Hilbert space   is the set \eq{
\tilde{\mathfrak{h}}=\big\{\vec{f}\in L^2(S_{\sigma,\Lambda};\mathbb{C}\otimes\mathbb{R}^3) \big|\forall \vec{k}\in S_{\sigma,\Lambda} \: a.e.:\:\vec{k}\cdot \vec{f}(\vec{k})=0\big\} 
}
of square-integrable, transversal vector fields, allowing momenta only in
\eq{
S_{\sigma,\Lambda}:=\big\{\vec{k}\in \mathbb{R}^3\big| \sigma \leq |\vec{k}|\leq \Lambda\big\}.
} 
The positive numbers $0<\sigma<\Lambda<\infty$ are called the infrared and ultraviolet cutoffs respectively.
For $\vec{k}\in S_{\sigma,\Lambda}$ the vectors $\vec{\varepsilon}_+(\vec{k})$ and $\vec{\varepsilon}_-(\vec{k})$ are chosen such that $\Big\{\vec{\varepsilon}_+(\vec{k}),\vec{\varepsilon}_-(\vec{k}), \frac{\vec{k}}{|\vec{k}|}\Big\}\subseteq \mathbb{R}^3$ is an orthonormal basis. Furthermore $\vec{k}\mapsto\vec{\varepsilon}_+(\vec{k})$ and $\vec{k}\mapsto\vec{\varepsilon}_-(\vec{k})$ are chosen to be measureable and in a way so that
\eq{
\vec{\varepsilon}_\pm(-\vec{k})=-\vec{\varepsilon}_\pm(\vec{k})
} holds true almost everywhere. $\tilde{\mathfrak{h}}$ may be identified with \eq{
\mathfrak{h}=L^2(S_{\sigma,\Lambda}\times \mathbb{Z}_2)
} through the unitary map
\eq{
\mathfrak{h} \ni f(\vec{k},\tau) \mapsto \vec{\varepsilon}_+ f(\vec{k},+) + \vec{\varepsilon}_- f(\vec{k},-) \in \tilde{\mathfrak{h}}.
}
In $\eqref{Pauli-Fierz Hamiltonian}$ the energy of the photon field is defined as the second quantization
\eq{ H_f = \mathrm{d}\Gamma\big(|k|\big),}
and the magnetic vector potential is given by
\eq{
\big(\vec{\mathbb{A}}(\vec{x})\big)_\nu
= a^\star\big(e^{-i\vec{k}\cdot \vec{x}}G_\nu(k)\big)+a\big(e^{-i\vec{k}\cdot \vec{x}}G_\nu(k)\big),
}
with $k= (\vec{k},\tau)$ and \eq{
\vec{G}(\vec{k},\tau):=g\:\vec{\varepsilon}_\tau(\vec{k})\:|\vec{k}|^{-\frac{1}{2}},
} for some constant $g\in \mathbb{R}$.
Moreover the total momentum operator $\vec{p}=-i\vec{\nabla} +\vec{P}_f$, where 
\eq{
\big(\vec{P}_f\big)_\nu=\big(\mathrm{d}\Gamma\big(\vec{k}\big)\big)_\nu=\mathrm{d}\Gamma(k_\nu)
} the photon field momentum,
commutes with the Pauli-Fierz Hamiltonian and hence a unitary transformation to the direct integral representation 
\eq{
\mathbb{U}\tilde{H}_g \mathbb{U}^\star = \int^\oplus H_{g,\vec{p}} \:\mathrm{d}^3p,
} is possible, where \eq{
H_{g,\vec{p}}=\frac{1}{2} \big(\vec{P}_f + \vec{\mathbb{A}}(\vec{0})-\vec{p}\big)^2+H_f.
}

\underline{References:} This presentation of the Pauli-Fierz Hamiltonian is taken from Chapter I of \cite{BBT}.

\subsection{Approximations of the Ground State Energy}

In this subsection the notion of ground state energy and its approximations studied in \cite{BBT} and \cite{BH} are introduced. Throughout this section we implicitly assume that density matrices are chosen in a way that their expectation values exist. To be more precise: Let $(A,\mathcal{D})\in\mathfrak{L}(\mathfrak{h})$. If we write $\Tr(\rho A)$ for a density matrix $\rho\in \mathfrak{DM}$, then we assume that $\rho A$ is a bounded operator that can be extented to all of $\mathfrak{h}$ and that this extension lies in $\mathcal{L}^1(\mathfrak{h})$. If we write $\la \varphi | A \varphi \ra$ for $\varphi\in\mathfrak{h}$, then $\varphi\in \mathcal{D}$ is assumed. 
\\
\\Given a Hamiltonian $H$ the \textbf{ground state energy $E_{gs}(H)$ of $H$} is the infimum of its spectrum
\eq{
E_{gs}(H):=\inf\{ \sigma(H)\}.
}
By virtue of the Rayleigh-Ritz principle the groundstate energy may be expressed as 
\eq{
E_{gs}(H)=\inf\big\{\la \psi | H\psi\ra \:|\: \psi\in \mathcal{D}(H),\|\psi\|=1\big\},
}
or due to the convexity of the density matrices even by
\eq{
E_{gs}(H)=\inf\big\{\Tr(\rho H)\: | \:\rho\in \mathcal{L}^1(\mathfrak{h}), \rho\geq 0, \Tr(\rho)=1 \big\}.
}
One would like to prove the existence of a \textbf{ground state} $\psi_{gs}\in \mathcal{D}(H)$, that is a normed vector satisfying $E_{gs}(H)=\la\psi_{gs}|H \psi_{gs}\ra$ or the existence of a density matrix $\rho_{gs}\in \mathfrak{DM}$ satisfying $E_{gs}(H)= \Tr(\rho H)$. Also one would like to identify the ground state vector and to calculate the numeric value of $E_{gs}(H)$. 
In order to approximate the ground state energy of the Pauli-Fierz Hamiltonian, the authors of \cite{BH} study the Lieb-Loss approximation
\eq{
E_{LL}\big(\tilde{H}_g\big):=\big\{ \mathcal{E}_{LL}(\phi_{el},\psi_{ph}) \:|\: \phi_{el}\in\mathfrak{h}_{el},\psi_{ph}\in\mathfrak{F}_{ph}, \|\phi_{el}\|=\|\psi_{ph}\|=1 \big\},
}
where
\eq{
\mathcal{E}_{LL}(\phi_{el},\psi_{ph}):= \big\la \phi_{el} \otimes \psi_{ph} \big| \tilde{H}_g (\phi_{el} \otimes \psi_{ph})\big\ra.
}
Obviously the Lieb-Loss approximation is an upper bound on the ground state energy \eq{E_{gs}(\tilde{H}_g)\leq E_{LL}(\tilde{H}_g),}
and asymtotics of $E_{LL}\big(\tilde{H}_g\big)$ with regards to $\Lambda\to \infty$ and $g\to 0$ are shown in [4, Theorem I.1]
\eq{
-C g^\frac{4}{49}  \Lambda^{-\frac{4}{49}}\leq \frac{E_{LL}\big(\tilde{H}_g\big)}{(\frac{4}{9 \pi}) F_1 (4\pi g)^{2/7} \Lambda^{12/7}}-1\leq C  g^\frac{4}{105}  \Lambda^{-\frac{4}{105}},
}
where $C>0$ is a universal constant and $F_1>0$ is the infimum of
\eq{\mathcal{F}_1(\phi):=\frac{1}{2}\big\|\vec{\nabla} \phi\big\|_2^2+ \|\phi\|_1,
} on the set $\phi\in H^1(\mathbb{R}^3)\cap L^1(\mathbb{R}^3)$ with $ \|\phi\|_2 =1$. Thus it is shown that $E_{gs}\big(\tilde{H}_g\big)$ grows at worst at the rate of $\Lambda^\frac{12}{7}$ asymtotically in $\Lambda\to \infty.$ The authors of \cite{BBT} study the Bogolubov-Hartree-Fock approximation of $H_{g,\vec{0}}$. 
The Bogolubov-Hartree-Fock energy is the infimum over quasifree density matrices
\eq{
E_{BHF}\big(H_{g,\vec{0}}\big)=\inf\big\{\Tr\big(\rho H_{g,\vec{0}}\big)\: | \:\rho\in \mathfrak{QF}\big\}.
}
In [3, Theorem IV.5] the authors show that the Bogolubov-Hartree-Fock energy coincides with the infimum over pure quasifree density matrices 
\eq{E_{BHF}\big(H_{g,\vec{0}}\big)=\inf \big\{ \big\la \mathbb{U}_B\mathbb{W}_\eta \Omega \big| H_{g,\vec{0}} \mathbb{W}_\eta^\star \mathbb{U}_B^\star\Omega \big\rangle \big| B\in \Bog, \eta \in \mathfrak{h} \big\},
} 
where the result [3, Lemma IV.1] on the appearance of pure quasifree density matrices is used. 
The existence of a minimizer in the set of pure quasifree density matrices is established in [3, Theorem VIII.3].

\underline{References:} The introduction of ground state energy is inspired by the chapter on Bogolubov-Hartee-Fock theory of \cite{HFT}.

\newpage
\section{Expectation Value of $H_{g,\vec{0}}$ in a Pure Quasifree State}
The goal is to express the expectation value $\big<\Omega \big|\mathbb{U}_B \mathbb{W}_\eta H_{g,\vec{0}} \mathbb{W}_\eta^\star \mathbb{U}_B^\star \Omega \big>$ of the fiber with vanishing total momentum $H_{g,\vec{0}}$ in a quasifree pure state as a functional of $U,V$ and $\eta$, where $B={\big(\begin{smallmatrix}
U & JVJ\\
V & JUJ
\end{smallmatrix}\big)}\in \Bog, \eta\in \mathfrak{h}$.
In Section III.1 the operator $H_{g,\vec{0}}$ is expressed in terms of the field operators $A_J,A^\star_J$, thereby preparing to compute the expectation of $H_{g,\vec{0}}$ for pure quasifree states in Section III.3 with the knowledge about expectations of field operators acquired in Section III.2.

\subsection{Expressing $H_{g,\vec{0}}$ through Field Operators}

In order to deal with the unbounded operator $H_{g,\vec{0}}$ the question of a suitable domain needs to be adressed. Since this is not the main focus of this work however, these considerations will only be sketched briefly. 
\\
\\First, we notice that each part of (II.66) is a sum of concatenations up to fourth order of bosonic creation and annihilation operators. Their natural domain is $P\mathcal{D}\big({\mathcal{N}^\frac{1}{2}}\big)$ and they are $\big(\mathcal{N}^\frac{1}{2},P\mathcal{D}\big(\mathcal{N}^\frac{1}{2}\big)\big)$-bounded, making them continuous as a map from $\big(P\mathcal{D}\big(\mathcal{N}^\frac{1}{2}\big),\\ \|\cdot\|_{\mathcal{N}^\frac{1}{2}}:=\|\cdot\|+\|\mathcal{N}^\frac{1}{2} \cdot \|\big)$ to $\mathfrak{F}_b(\mathfrak{h})$. Since $\mathfrak{F}_b(\mathfrak{h})$ is a closed subspace of $\mathfrak{F}(\mathfrak{h})$ it inherits the Hilbert space property making $\mathfrak{B}\big(P\mathcal{D}\big(\mathcal{N}^\frac{1}{2}\big);\mathfrak{F}_b(\mathfrak{h}
)\big)$ a Banach space. This way it is possible to make sense of limits of unbounded operators on $\mathfrak{h}$ as long as they are $\mathcal{N}^\frac{1}{2}$-bounded and one may show identities such as \eq{
A^\star_J(F)=\sum\nolimits_{i=1}^\infty \la F_i|F\ra A^\star_J(F_i)
}
where $\{F_i\}_i\subseteq \mathfrak{h}\oplus\mathfrak{h}$ is an ONB and $F\in\mathfrak{h}\oplus\mathfrak{h}$.
\\ Secondly, we make the observation that the creation and annihilation operators act similarly to $\mathcal{N}^\frac{1}{2}$ in the sense that they map $P\mathcal{D}\big(\mathcal{N}^\frac{k}{2}\big)$ into $P\mathcal{D}\big(\mathcal{N}^\frac{k-1}{2}\big)$ for any $k\in\mathbb{N}$. This suggests that the dense subspace $P\mathcal{D}\big(\mathcal{N}^2\big)\subseteq \mathfrak{F}_b(\mathfrak{h})$ is a suitable domain for $H_{g,\vec{0}}$. 
\\ Lastly, we can see that expressions like (III.1) and second quantizations (II.43) of bounded operators on $\mathfrak{h}$ also converge in $\mathfrak{B}\big(\mathcal{D}\big(\mathcal{N}^2\big);\mathfrak{F}_b(\mathfrak{h})\big)$.

\Lemma{Let $H_f,\vec{\mathbb{A}}\big(\vec{0}\big)$ and $\vec{P}_f$ be defined as in section II.8.
Then the following equalities hold true.
\begin{enumerate}[label=(\roman*)]
\item $H_{f} =  \mathrm{d}\Gamma_J\big[{\big(\begin{smallmatrix}
|k| & 0 \\
0 & 0 
\end{smallmatrix}\big)},0\big],$
\item $\vec{\mathbb{A}}\big(\vec{0}\big)^2 =\sum_{\nu=1}^3 \mathrm{d} \Gamma_J\big[| q(G_\nu) \rangle \langle q(G_\nu)|,0\big],$
\item $\vec{P}_f^2= \sum_{\nu=1}^3 \mathrm{d}\Gamma_J\big[{\big(\begin{smallmatrix}
k_\nu & 0 \\
0 & 0 
\end{smallmatrix}\big)},0\big]^2,$
\item $\vec{P}_f \vec{\mathbb{A}}\big(\vec{0}\big) +\vec{\mathbb{A}}\big(\vec{0}\big)\vec{P}_f =\sum_{\nu=1}^3 \Big\{ \mathrm{d}\Gamma_J\big[{\big(\begin{smallmatrix}
k_\nu & 0 \\
0 & 0 
\end{smallmatrix}\big)},0\big] A_J\big(q(G_\nu)\big) +A_J\big(q(G_\nu)\big)\mathrm{d}\Gamma_J\big[{\big(\begin{smallmatrix}
k_\nu & 0 \\
0 & 0 
\end{smallmatrix}\big)},0\big]  \Big\}.$
\end{enumerate}
}
\begin{proof}
Let $\{\varphi_i\}_{i\in\mathbb{N}}$ be an orthonormal basis of $\mathfrak{h}$ and let $\{F_i\}_{i\in \mathbb{N}}$ be an orthonormal basis of $\mathfrak{h}\oplus \mathfrak{h}$. Note that the set $\{(\varphi_i\oplus 0), (0\oplus \varphi_i)\}_{i\in\mathbb{N}}$ is a possible choice for an orthonormal basis of $\mathfrak{h}\oplus \mathfrak{h}$ and that the second quantization $\mathrm{d}\Gamma_J$ is independent of the choice of basis. Moreover, we observe that multiplication by $|k|$ and $k_\nu$ define bounded operators on $\mathfrak{h}=L^2(S_{\sigma,\Lambda}\times\mathbb{Z}_2)$, thereby justifying us in treating the above operators as bounded operators in $\mathfrak{B}\big(\mathcal{D}\big(\mathcal{N}^2\big);\mathfrak{F}_b(\mathfrak{h})\big)$.
\\$(i)$ 
The energy of the photon field $H_f$ can easily be rewritten from a second quantization $\mathrm{d}\Gamma$ into a second quantization $\mathrm{d}\Gamma_J$
\begin{align}
\notag H_{ph} &= \mathrm{d}\Gamma(|k|) 
= \sum\nolimits_{i,j = 1}^\infty \big< \varphi_i \big||k|\varphi_j \big> a^\star(\varphi_i) a(\varphi_j) \\&= \sum\nolimits_{i,j =1}^\infty \big< F_i \big| 
\big(\begin{smallmatrix}
|k| & 0 \\
0 & 0 
\end{smallmatrix}\big) F_j \big> A^\star_J(F_i) A_J(F_j)= \mathrm{d}\Gamma_J\big[{\big(\begin{smallmatrix}
|k| & 0 \\
0 & 0 
\end{smallmatrix}\big)},0 \big],
\end{align}
due to $\big\la 0\oplus \varphi_i \big|\big(\begin{smallmatrix} |k| &0\\0&0 \end{smallmatrix}\big) \varphi_j \oplus 0 \big\ra=\big\la \varphi_i\oplus 0 |\big(\begin{smallmatrix} |k| &0\\0&0 \end{smallmatrix}\big) 0 \oplus \varphi_j \big\ra=\big\la 0\oplus \varphi_i |\big(\begin{smallmatrix} |k| &0\\0&0 \end{smallmatrix}\big) 0\oplus \varphi_j \big\ra=0$ for $i,j\in\mathbb{N}$.
\\$(ii)$
By using (III.1) the square of the magnetic vector potential becomes

\begin{align}
\notag \vec{\mathbb{A}}\big(\vec{0}\big)^2  
&= \sum\nolimits_{\nu=1}^3 \big(a^\star(G_\nu)+a(G_\nu) \big)^2 =\sum\nolimits_{\nu=1}^3 A_J^\star\big(q(G_\nu)\big) A_J \big(q(G_\nu)\big) \\ \notag &= \sum\nolimits_{\nu=1}^3 \sum\nolimits_{i,j=1}^\infty \big\langle F_i \big| q(G_\nu) \big\rangle \big\langle q(G_\nu) \big| F_j \big\rangle A_J^\star(F_i)A_J(F_j) \\&= \sum\nolimits_{\nu=1}^3 \mathrm{d} \Gamma_J\big[| q(G_\nu) \rangle \langle q(G_\nu)|,0\big], 
\end{align}
\\$(iii)$
Similarly to (III.2) the photon field momentum can be rewritten as

\begin{align}
\notag \vec{P}^2_f =\sum\nolimits_{\nu=1}^3 \big(\mathrm{d}\Gamma(k_\nu)\big)^2 &=\sum\nolimits_{\nu=1}^3 \left(\sum\nolimits_{i,j=1}^\infty \left\langle \varphi_i \middle| k_\nu\varphi_j \right\rangle a^\star(\varphi_i)a(\varphi_j) \right)^2 
\\ &= \sum\nolimits_{\nu=1}^3 \mathrm{d}\Gamma_J\big[{\big(\begin{smallmatrix}
k_\nu & 0 \\
0 & 0 
\end{smallmatrix}\big)},0\big]^2.
\end{align}
\\$(iv)$
The cross terms consequently assume the form
\begin{align}
\notag &\vec{P}_f \vec{\mathbb{A}}\big(\vec{0}\big) +\vec{\mathbb{A}}\big(\vec{0}\big)\vec{P}_f \\=& \sum\nolimits_{\nu=1}^3 \Big\{ \mathrm{d}\Gamma_J\big[\big(\begin{smallmatrix}
k_\nu & 0 \\
0 & 0 
\end{smallmatrix}\big),0\big] A_J\big(q(G_\nu)\big) +A_J\big(q(G_\nu)\big)\mathrm{d}\Gamma_J\big[\big(\begin{smallmatrix}
k_\nu & 0 \\
0 & 0 
\end{smallmatrix}\big),0\big]  \Big\}.
\end{align}
\end{proof}

\noindent As a corollary we can conclude
\Theorem{The fiber of total momentum $\vec{p}=\vec{0}$ of the Pauli-Fierz Hamiltonian fulfills the following equation
\begin{align}
\notag H_{g,\vec{0}} =& \frac{1}{2} \sum\nolimits_{\nu=1}^3 \Big\{ 
\mathrm{d}\Gamma_J\big[\big(\begin{smallmatrix}
k_\nu & 0 \\
0 & 0 
\end{smallmatrix}\big),0\big]^2 
+ \mathrm{d} \Gamma_J\big[| q(G_\nu) \rangle \langle q(G_\nu)|,0\big] 
+\mathrm{d}\Gamma_J\big[\big(\begin{smallmatrix}
k_\nu & 0 \\
0 & 0 
\end{smallmatrix}\big),0\big] \\ & \cdot A_J\big(q(G_\nu)\big) 
+ A_J\big(q(G_\nu)\big)\mathrm{d}\Gamma_J\big[{\big(\begin{smallmatrix}
k_\nu & 0 \\
0 & 0 
\end{smallmatrix}\big)},0 \big] 
\Big\} 
+ \mathrm{d}\Gamma\big[{\big(\begin{smallmatrix}
|k| & 0 \\
0 & 0 
\end{smallmatrix}\big)},0\big] \label{H in zweiter Quantisierung}.
\end{align}
}

\subsection{Expectation Values of Strings of Field Operators}
Since the highest order term in $\eqref{H in zweiter Quantisierung}$ is $ \mathrm{d}\Gamma\big[\big(\begin{smallmatrix}
k_\nu & 0\\
0 & 0
\end{smallmatrix}\big),0\big]^2$, 
the next step is to calculate expectation values of strings of field operators up to the fourth order in a pure quasifree state.
If $F \in \mathfrak{h}\oplus \mathfrak{h}$ we will use the short hand notation
\eq{
\forall \eta\in\mathfrak{h}: Q_\eta(F) &:= \big\langle q(\eta) \big| F \big\rangle
}
and if $\big(T,P\mathcal{D}\big(\mathcal{N}^2\big)\big)\in \mathfrak{L}\big(\mathfrak{F}_b(\mathfrak{h})\big)$ we will make the abbreviation
\eq{
\forall B\in\Bog,\forall\eta \in \mathfrak{h} : \:&E_{\eta,B}(T):= \big<\Omega \big|\mathbb{U}_B \mathbb{W}_\eta T \mathbb{W}_\eta^\star \mathbb{U}_B^\star \Omega \big>.
}
\Bem{
Note that for (III.8) to be defined we implicitly assume\eq{
\mathbb{W}^\star_\eta &\mathbb{U}^\star_B \Omega\in P\mathcal{D}\big(\mathcal{N}^2\big).
} 
}

\Theorem{\label{Expectation of strings of annihilators}
Let $\big( \begin{smallmatrix} 
U & JVJ\\V& JUJ \end{smallmatrix} \big)\in \Bog$ and $\eta \in \mathfrak{h}$. Then for $F_1,F_2,F_3, F_4 \in \mathfrak{h} \oplus \mathfrak{h}$ the following identities hold
\begin{align}
E_{\eta,B}\big(A_J(F_1)\big)=& \overline{Q_\eta(F_1)},
\\ E_{\eta,B}\big(A_J(F_1)A_J(F_2)\big)=& \big\langle F_1 \big| B^\star M B F_2 \big\rangle +\overline{Q_\eta(F_1)}\overline{Q_\eta(F_2)},
\\ \notag E_{\eta,B}\Big(\prod\nolimits_{k=1}^3 A_J(F_k)\Big)=& \big\langle F_1 \big| B^\star M B F_2 \big\rangle\: \overline{Q_\eta(F_3)}+\big\langle F_1 \big| B^\star M B F_3 \big\rangle\: \overline{Q_\eta(F_2)} \\&+\big\langle F_2 \big| B^\star M B F_3 \big\rangle \:\overline{Q_\eta(F_1)}+ \prod\nolimits_{k=1}^3 \overline{Q_\eta(F_k)},
\\ \notag E_{\eta,B}\Big(\prod\nolimits_{k=1}^4 A_J(F_k)\Big)=& \sum\nolimits_{\sigma \in P_{4}} \big\langle F_{\sigma(1)} \big| B^\star M B F_{\sigma(2)} \big\rangle \big\langle F_{\sigma(3)} \big| B^\star M B F_{\sigma(4)} \big\rangle \\ \notag &+ \sum\nolimits_{\tau\in X} \big\langle F_{\tau(1)} \big| B^\star M B F_{\tau(2)} \big\rangle \: \overline{Q_\eta(F_{\tau(3)})}\overline{Q_\eta(F_{\tau(4)})} \\&+ \prod\nolimits_{k=1}^4 \overline{Q_\eta(F_k)},
\end{align}
where $P_{4}=\{ \sigma \in \mathcal{S}_4 : \sigma(1)<\sigma(2), \sigma(3)<\sigma(4),\sigma(1)<\sigma(3)\}$ is the set of pairings, $X:=\{ \tau \in \mathcal{S}_4 : \tau(1)<\tau(2), \tau(3)<\tau(4)\}$ and \eq{M := \big(\begin{smallmatrix}
0 & J \\
0 & 0 
\end{smallmatrix}\big).}
}

\begin{proof}
We will freely use the Wick Theorem~II.10 throughout this proof. Let $\Psi:=\mathbb{U}_B^\star \Omega$ and $F_k = f_k \oplus g_k \in \mathfrak{h} \oplus \mathfrak{h}$ for $k=1,2,3,4$. Since $\langle \Psi |A_J(F_1)\Psi\rangle=0$ and (II.48) we swiftly conclude
\begin{align}
E_{\eta,B}\big(A_J(F_1)\big)=\big\langle \Psi \big|\big(A_J(F_1)+\overline{Q_\eta(F_1)}\big) \Psi\big\rangle =\overline{Q_\eta(F_1)}.
\end{align}
For the second order we can see that
\begin{align}
\notag &\big\langle \Psi \big|A_J(F_1)A_J(F_2)\Psi\big\rangle=\big\langle \Omega \big|A_J(BF_1)A_J(BF_2)\Omega\big\rangle 
\\ \notag =&\big\langle \Omega \big|a(Uf_1+JVJg_1)a^\star(JVf_2+UJg_2)\Omega\big\rangle = \langle Uf_1+JVJg_1 | JVf_2+UJg_2 \rangle 
\\ \notag =& \langle f_1| U^\star JV f_2 \rangle + \langle f_1 | U^\star UJ g_2 \rangle + \langle g_1 |JV^\star Vf_2 \rangle + \langle g_1 | JV^\star JUJg_2 \rangle \\=& \big\langle F_1 \big| \big(\begin{smallmatrix}
U^\star JV & U^\star U\\
JV^\star V & JV^\star JUJ
\end{smallmatrix}\big) F_2 \big\rangle=\big\langle F_1 \big| B^\star MB F_2 \big\rangle,
\end{align}
which leads us to
\begin{align}
\notag E_{\eta,B}\big(A_J(F_1)A_J(F_2)\big)&= \big\langle \Psi \big| \big(A_J(F_1)+ \overline{Q_\eta(F_1)}\big)\big(A_J(F_2)+ \overline{Q_\eta(F_2)}\big) \Psi\big\rangle \\&= \big\langle F_1 \big| B^\star MB F_2 \big\rangle + \overline{Q_\eta(F_1)}\overline{Q_\eta(F_2)}.
\end{align}
The third order again vanishes $\big\langle \Psi \big|A_J(F_1)A_J(F_2)A_J(F_3)\Psi\big\rangle =0$, and we easily see
\begin{align}
\notag E_{\eta,B}\Big(\prod\nolimits_{k=1}^3 A_J(F_k)\Big)=& \Big\langle \Psi \Big|\prod\nolimits_{k=1}^3 \big(A_J(F_k)+ \overline{Q_\eta(F_k)}\big)\Psi\Big\rangle 
\\ \notag =& \big\langle F_1 \big| B^\star MB F_2 \big\rangle\: \overline{Q_\eta(F_3)}
+\big\langle F_1 \big| B^\star MB F_3 \big\rangle\: \overline{Q_\eta(F_2)}
\\&+\big\langle F_2 \big| B^\star MB F_3 \big\rangle\: \overline{Q_\eta(F_1)} +\prod\nolimits_{k=1}^3 \overline{Q_\eta(F_k)}.
\end{align}
Finally, the forth order expectation computes to
\begin{align}
\notag &E_{\eta,B}\Big(\prod\nolimits_{k=1}^4 A_J(F_k)\Big)= \Big\langle \Psi \Big|\prod\nolimits_{k=1}^4 \big(A_J(F_k)+ \overline{Q_\eta(F_k)}\big)\Psi\Big\rangle 
\\ \notag =& \Big\langle \Psi \Big|\prod\nolimits_{k=1}^4 A_J(F_k) \Psi\Big\rangle  + \prod\nolimits_{k=1}^4 \overline{Q_\eta(F_k)}
\\ \notag &+\sum\nolimits_{\tau\in X} \big\langle F_{\tau(1)} \big| B^\star M B F_{\tau(2)} \big\rangle \:\overline{Q_\eta(F_{\tau(3)})}\overline{Q_\eta(F_{\tau(4)})} 
\\ \notag =& \sum\nolimits_{\sigma \in P_{4}} \big\langle F_{\sigma(1)} \big| B^\star M B F_{\sigma(2)} \big\rangle \big\langle F_{\sigma(3)} \big| B^\star M B F_{\sigma(4)} \big\rangle + \prod\nolimits_{k=1}^4 \overline{Q_\eta(F_k)}\\&+\sum\nolimits_{\tau\in X} \big\langle F_{\tau(1)} \big| B^\star M B F_{\tau(2)} \big\rangle\: \overline{Q_\eta(F_{\tau(3)})}\overline{Q_\eta(F_{\tau(4)})} .
\end{align}
\end{proof}

\subsection{Expectation Value of $H_{g,\vec{0}}$}
We are now in a position to calculate the expectiation value $E_{\eta,B}(H_{g,\vec{0}})$ of $H_{g,\vec{0}}$ in a pure quasifree state.

\Theorem{
Let $B=\big(\begin{smallmatrix} U & JVJ\\ V & JUJ \end{smallmatrix}\big)\in\Bog$ and $\eta \in \mathfrak{h}$. Then
\begin{align}
\notag E_{\eta,B}\big( H_{g,\vec{0}} \big) =&
\frac{1}{2} \sum\nolimits_{\nu=1}^3 \Big\{ 
\big(\mathrm{Tr}[k_\nu V^\star V ]+\langle \eta | k_\nu \eta \rangle +2 \mathrm{Re}\langle \eta | G_\nu \rangle \big)^2 
\\ \notag &+ \mathrm{Tr}\big[(V^\star JU k_\nu)^2\big]
+ \mathrm{Tr} [k_\nu V^\star V k_\nu (1+V^\star V)]
\\ \notag &+ \langle G_\nu +k_\nu \eta | (1+ 2V^\star V) (G_\nu +k_\nu \eta) \rangle
\\ 
&+ 2\mathrm{Re}\langle G_\nu + k_\nu\eta | V^\star JU (G_\nu +k_\nu \eta) \rangle
\Big\} +\mathrm{Tr}[|k|V^\star V]+\big\langle \eta \big| |k| \eta \big\rangle,
\end{align}
where $E_{\eta,B}\big( H_{g,\vec{0}} \big) =\big<\Omega \big|\mathbb{U}_B \mathbb{W}_\eta H_{g,0} \mathbb{W}_\eta^\star \mathbb{U}_B^\star \Omega \big>$.
}
\begin{proof}
We separately calculate the expectation of every term of $\eqref{H in zweiter Quantisierung}$. For the convenience of the reader we list the following matrix identities, which we will be making use of extensively.
\begin{align}
\notag B=\begin{pmatrix}
U & JVJ\\
V & JUJ
\end{pmatrix} \:\:\:
&  B\mathcal{J} = \begin{pmatrix}
JV & UJ\\
JU &VJ
\end{pmatrix} \\
MB =\begin{pmatrix}
JV & UJ\\
0 & 0
\end{pmatrix} \:\:\:& 
MB\mathcal{J} = \begin{pmatrix}
U & JVJ\\
0 & 0
\end{pmatrix}
\end{align}
We will also employ the conditions $U^\star U = 1 + V^\star V$ and $U^\star JV = V^\star JU$ in order to simplify the resulting expressions. Moreover, we observe that 
\begin{align}
\forall F \in \mathfrak{h}\oplus \mathfrak{h}: \overline{Q_\eta(\mathcal{J}F)}=Q_\eta(F),\\ \forall F \in \mathfrak{h}\oplus \mathfrak{h}: A^\star(F)=A(\mathcal{J}F).
\end{align}
Let $\{\varphi_i\}_{i\in \mathbb{N}} $ be an orthonormal basis of $\mathfrak{h}$ and let $F_i:= \varphi_i \oplus 0$ for $o\in \mathbb{N}$. Then using Theorem~\ref{Expectation of strings of annihilators} we can calculate the expectation value of the squared photon field momentum
\eq{
\notag &E_{\eta,B}\big( \mathrm{d}\Gamma_J\big[\big(\begin{smallmatrix}
k_\nu & 0 \\
0 & 0 
\end{smallmatrix}\big),0\big]^2 \big) 
\\ \notag =&
\sum\nolimits_{i,j,k,l=1}^\infty \big\langle F_i \big| \big(\begin{smallmatrix}
k_\nu & 0 
\\ 0 & 0
\end{smallmatrix}\big) F_j \big\rangle \big\langle F_k \big| \big(\begin{smallmatrix}
k_\nu & 0 
\\ 0 & 0
\end{smallmatrix}\big) F_l \big\rangle 
E_{\eta,B}\big(A_J(\mathcal{J}F_i)A_J(F_j)A_J(\mathcal{J}F_k)A_J(F_l) \big)\\
\notag =& \sum\nolimits_{i,j,k,l=1}^\infty \big\langle F_i \big| \big(\begin{smallmatrix}
k_\nu & 0 
\\ 0 & 0
\end{smallmatrix}\big) F_j \big\rangle \big\langle F_k \big| \big(\begin{smallmatrix}
k_\nu & 0 
\\ 0 & 0
\end{smallmatrix}\big) F_l \big\rangle \Big\{\big\langle \mathcal{J} F_i \big| B^\star MB F_j \big\rangle \big\langle \mathcal{J} F_k \big| B^\star MB F_l \big\rangle 
\\ \notag &+ \big\langle \mathcal{J} F_i \big| B^\star MB \mathcal{J}F_k \big\rangle \big\langle F_j \big| B^\star MB F_l \big\rangle +
\big\langle \mathcal{J} F_i \big| B^\star MB F_l \big\rangle \big\langle F_j \big| B^\star MB \mathcal{J}F_k \big\rangle\\
\notag  &+
\big\langle \mathcal{J} F_i \big| B^\star MB F_j \big\rangle \:
\overline{Q_\eta(\mathcal{J}F_k)} 
\overline{Q_\eta(F_l)}+
\big\langle \mathcal{J} F_i \big| B^\star MB \mathcal{J} F_k \big\rangle \:
\overline{Q_\eta(F_j)} 
\overline{Q_\eta(F_l)}\\
\notag &+
\big\langle \mathcal{J} F_i \big| B^\star MB F_l \big\rangle \:
\overline{Q_\eta(F_j)} 
\overline{Q_\eta(\mathcal{J} F_k)}+
\big\langle  F_j \big| B^\star MB \mathcal{J}F_k \big\rangle \:
\overline{Q_\eta(\mathcal{J}F_i)} 
\overline{Q_\eta(F_l)} 
}
\eq{\notag &+
\big\langle  F_j \big| B^\star MB F_l \big\rangle \:
\overline{Q_\eta(\mathcal{J}F_i)} 
\overline{Q_\eta(\mathcal{J}F_k)}+
\big\langle \mathcal{J} F_k \big| B^\star MB F_l \big\rangle \:
\overline{Q_\eta(\mathcal{J}F_i)} 
\overline{Q_\eta(F_j)}
\\ \notag &+ \overline{Q_\eta(\mathcal{J}F_i)} \overline{Q_\eta(F_j)} \overline{Q_\eta(\mathcal{J}F_k)} \overline{Q_\eta(F_l)} \Big\}
\\ \notag =& \sum\nolimits_{i,j,k,l=1}^\infty \langle \varphi_i | k_\nu \varphi_j \rangle \langle \varphi_k | k_\nu \varphi_l \rangle \Big\{
\langle \varphi_j | V^\star V \varphi_i \rangle \langle \varphi_l | V^\star V \varphi_k \rangle
+ \langle U^\star JV \varphi_i | \varphi_k \rangle \langle \varphi_j | U^\star JV \varphi_l \rangle
\\ \notag & +\langle \varphi_l | V^\star V \varphi_i \rangle \langle \varphi_j | U^\star U \varphi_k \rangle
+ \langle \varphi_j | V^\star V \varphi_i \rangle \langle \eta | \varphi_k \rangle \langle \varphi_l |  \eta \rangle
+ \langle U^\star JV\varphi_i |  \varphi_k \rangle \langle \varphi_j |  \eta \rangle \langle \varphi_l |  \eta \rangle
\\ \notag &+ \langle \varphi_l | V^\star V \varphi_i \rangle \langle \varphi_j | \eta \rangle \langle \eta |  \varphi_k \rangle
+\langle \varphi_j | U^\star U \varphi_k \rangle \langle \eta |  \varphi_i \rangle \langle \varphi_l |  \eta \rangle
+\langle \varphi_j | U^\star JV \varphi_l \rangle \langle \eta |  \varphi_i \rangle \langle \eta | \varphi_k \rangle
\\ \notag &+ \langle \varphi_l | V^\star V \varphi_k \rangle \langle \eta |  \varphi_i \rangle \langle \varphi_j |  \eta \rangle \Big\}
\\ \notag =& \mathrm{Tr}[k_\nu V^\star V ]^2 
+ \sum\nolimits_{i,l=1}^\infty \langle U^\star JV \varphi_i | k_\nu \varphi_l \rangle \langle \varphi_i | k_\nu U^\star JV \varphi_l \rangle
+ \mathrm{Tr} [k_\nu V^\star V k_\nu U^\star U]
\\ \notag &+ \langle \eta | k_\nu \eta \rangle ^2
+\langle \eta | k_\nu \eta \rangle \mathrm{Tr}[k_\nu V^\star V]
+\sum\nolimits_{i=1}^\infty \langle U^\star JV \varphi_i |k_\nu \eta \rangle \langle \varphi_i | k_\nu \eta \rangle
+\langle \eta | k_\nu V^\star V k_\nu \eta \rangle
\\ \notag &+ \langle \eta | k_\nu U^\star U k_\nu \eta \rangle 
+\sum\nolimits_{l=1}^\infty \langle \eta | k_\nu U^\star JV \varphi_l \rangle \langle \eta | k_\nu \varphi_l \rangle
+\langle \eta | k_\nu \eta \rangle \mathrm{Tr}[k_\nu  V^\star V]
\\ \notag =& \mathrm{Tr}[k_\nu V^\star V ]^2 
+ \sum\nolimits_{l=1}^\infty \langle V^\star JU k_\nu \varphi_l | k_\nu U^\star JV \varphi_l \rangle
+ \mathrm{Tr} [k_\nu V^\star V k_\nu U^\star U]
\\ \notag &+ \langle \eta | k_\nu \eta \rangle ^2
+\langle \eta | k_\nu \eta \rangle \mathrm{Tr}[k_\nu V^\star V]
+ \langle V^\star JU k_\nu \eta |k_\nu \eta \rangle
+\langle \eta | k_\nu V^\star V k_\nu \eta \rangle
\\ \notag &+ \langle \eta | k_\nu U^\star U k_\nu \eta \rangle 
+ \langle \eta | k_\nu V^\star JU k_\nu \eta \rangle 
+\langle \eta | k_\nu \eta \rangle \mathrm{Tr}[k_\nu  V^\star V]
\\ \notag =& \big(\mathrm{Tr}[k_\nu V^\star V ]+\langle \eta | k_\nu \eta \rangle \big)^2 
+ \mathrm{Tr}\big[(V^\star JU k_\nu)^2\big]
+ \mathrm{Tr} \big[(k_\nu V^\star V)^2\big] +\mathrm{Tr}\big[k_\nu^2 V^\star V\big]
\\&+ 2\langle \eta | k_\nu V^\star V k_\nu \eta \rangle
+\langle \eta | k_\nu^2 \eta \rangle 
+ 2\mathrm{Re}\langle \eta | k_\nu V^\star JU k_\nu \eta \rangle,
}
where we used that $\langle U^\star JV \varphi_i | k_\nu \varphi_l \rangle = \langle V^\star JU k_\nu \varphi_l | \varphi_i \rangle$ and similar identities which are derived from the property of $J$ that for any $x,y\in \mathfrak{h}: \langle x | Jy \rangle = \langle y | Jx \rangle$. The other expectation values can be computed in a similar fashion. Let $\overline{M}:= \big(\begin{smallmatrix}
0 & 0
\\J &0
\end{smallmatrix}\big)$ be the matrix that fulfils $\langle x| My \rangle =\big\langle y \big| \overline{M}x \big\rangle, \forall x,y \in \mathfrak{h}$. Then the expectation of the squared magnetic vector potential is
\eq{
\notag &E_{\eta,B}\big(\mathrm{d} \Gamma_J\big[| q(G_\nu) \rangle \langle q(G_\nu)|,0\big]\big) \\
\notag =&  \sum\nolimits_{i,j=1}^\infty \langle F_i | q(G_\nu) \rangle \langle q(G_\nu) | F_j \rangle \: E_{\eta,B}\big(A_J(\mathcal{J}F_i)A_J(F_j)\big)
\\
\notag =& \sum\nolimits_{i,j=1}^\infty \langle F_i | q(G_\nu) \rangle \langle q(G_\nu) | F_j \rangle \: \big( \langle \mathcal{J} F_i | B^\star MB F_j \rangle +Q_\eta(F_i) \overline{Q_\eta(F_j)}\big)\\
\notag =& \big\langle q(G_\nu) \big| B^\star \overline{M} BJ q(G_\nu) \big\rangle+
\langle q(\eta) | q(G_\nu) \rangle \langle q(G_\nu) | q(\eta) \rangle 
\\
\notag =& \langle BJ q(G_\nu) | MB q(G_\nu) \rangle+
\big(\langle \eta | G_\nu \rangle + \langle G_\nu | \eta \rangle \big)^2
\\ \notag =& \langle (JV+U)G_\nu | (JV+U)G_\nu \rangle +4 \big(\mathrm{Re}\langle \eta | G_\nu\rangle \big)^2
\\ \notag =& \langle G_\nu | (V^\star V +U^\star U) G_\nu \rangle + 2\mathrm{Re} \langle G_\nu | V^\star JU G_\nu \rangle +4 \big(\mathrm{Re}\langle \eta | G_\nu\rangle \big)^2
\\ =& 2\langle G_\nu | V^\star V G_\nu \rangle + \|G_\nu\|^2 + 2\mathrm{Re} \langle G_\nu | V^\star JU G_\nu \rangle +4 \big(\mathrm{Re}\langle \eta | G_\nu\rangle \big)^2.
}
\newpage
Similiarly we can calculate the expectation of the cross terms

\begin{align}
\notag &E_{\eta,B}\Big(\mathrm{d}\Gamma_J\big[\big(\begin{smallmatrix}
k_\nu & 0 \\
0 & 0 
\end{smallmatrix}\big),0\big] A_J\big(q(G_\nu)\big) \Big)
\\ \notag =& \sum\nolimits_{i,j=1}^\infty \big\langle F_i \big| \big( \begin{smallmatrix}
k_\nu &0
\\0&0
\end{smallmatrix}\big) F_j\big\rangle \:E_{\eta,B}\big(A_J(\mathcal{J}F_i)A_J(F_j)A_J(q(G_\nu))\big)
\\ \notag =& \sum\nolimits_{i,j=1}^\infty \big\langle F_i \big| \big( \begin{smallmatrix}
k_\nu &0
\\0&0
\end{smallmatrix}\big) F_j\big\rangle \Big\{Q_\eta(F_i)\overline{Q_\eta(F_j)}\overline{Q_\eta(q(G_\nu))}
+ \langle JF_i |B^\star MB F_j\rangle \:\overline{Q_\eta(q(G_\nu))}
\\ \notag &+\langle JF_i| B^\star MB q(G_\nu)\rangle\: \overline{Q_\eta(F_j)} 
+\langle F_j| B^\star MB q(G_\nu)\rangle\: Q_\eta(F_i)\Big\} 
\\ \notag =& \big \langle q(\eta) \big| \big( \begin{smallmatrix}
k_\nu &0
\\0&0
\end{smallmatrix}\big) q(\eta) \rangle \langle q(G_\nu)|q(\eta)\rangle 
+ \sum\nolimits_{i,j=1}^\infty \langle\varphi_i| k_\nu \varphi_j \rangle \Big\{ \langle JV\varphi_i| JV \varphi_j \rangle \langle q(G_\nu)|q(\eta)\rangle 
\\ \notag &+   \langle JV\varphi_i| (JV+U) G_\nu \rangle \langle \varphi_j|\eta\rangle 
+  \langle U\varphi_j| (JV+U) G_\nu \rangle \langle \eta|\varphi_i \rangle \Big\}
\\ \notag =& \langle \eta | k_\nu \eta \rangle 2\mathrm{Re}\langle\eta|G_\nu\rangle +\mathrm{Tr}[k_\nu V^\star V] 2\mathrm{Re}\langle\eta|G_\nu\rangle +\langle k_\nu (V^\star JU +V^\star V) G_\nu |\eta\rangle \\
\notag &+\langle\eta |k_\nu(U^\star JV +U^\star U) G_\nu \rangle
\\ \notag =& \langle \eta | k_\nu \eta \rangle 2\mathrm{Re}\langle\eta|G_\nu\rangle +\mathrm{Tr}[k_\nu V^\star V] 2\mathrm{Re}\langle\eta|G_\nu\rangle + 2\mathrm{Re} \langle \eta | k_\nu U^\star JV G_\nu \rangle 
\\ &+ 2\mathrm{Re} \langle \eta | k_\nu V^\star V G_\nu \rangle + \langle \eta | k_\nu G_\nu\rangle,
\end{align}
and 
\begin{align}
\notag &E_{\eta,B}\Big(A_J\big(q(G_\nu)\big) \mathrm{d}\Gamma_J\big[\big(\begin{smallmatrix}
k_\nu & 0 \\
0 & 0 
\end{smallmatrix}\big),0\big] \Big)
\\ \notag =& \sum\nolimits_{i,j=1}^\infty \big\langle F_i \big| \big( \begin{smallmatrix}
k_\nu &0
\\0&0
\end{smallmatrix}\big) F_j\big\rangle \:E_{\eta,B}\big(A_J(q(G_\nu))A_J(\mathcal{J}F_i)A_J(F_j)\big)
\\ \notag =& \sum\nolimits_{i,j=1}^\infty \big\langle F_i \big| \big( \begin{smallmatrix}
k_\nu &0
\\0&0
\end{smallmatrix}\big) F_j\big\rangle \Big\{\overline{Q_\eta(q(G_\nu))}Q_\eta(F_i)\overline{Q_\eta(F_j)}
+ \langle q(G_\nu) |B^\star MB \mathcal{J} F_i\rangle\: \overline{Q_\eta(F_j)}
\\ \notag &+\langle q(G_\nu)| B^\star MB F_j\rangle \:Q_\eta(F_i) 
+\langle \mathcal{J} F_i| B^\star MB F_j \rangle \:\overline{Q_\eta(q(G_\nu))}\Big\}
\\ \notag =&  \langle \eta | k_\nu \eta \rangle 2\mathrm{Re}\langle \eta|G_\nu\rangle 
+ \sum\nolimits_{i,j=1}^\infty \langle\varphi_i| k_\nu \varphi_j \rangle 
\Big\{ 
\langle (U+JV)G_\nu | U \varphi_i \rangle \langle \varphi_j|\eta\rangle 
\\ \notag &+   \langle (U+JV) G_\nu| JV \varphi_j \rangle \langle \eta|\varphi_i \rangle 
+  \langle JV \varphi_i| JV \varphi_j \rangle 2\mathrm{Re}\langle \eta|G_\nu \rangle 
\Big\}
\\ \notag =&  \langle \eta | k_\nu \eta \rangle 2\mathrm{Re}\langle \eta|G_\nu\rangle  + \mathrm{Tr}[k_\nu V^\star V] 2\mathrm{Re}\langle\eta|G_\nu\rangle + \langle k_\nu (U^\star U +U^\star JV) G_\nu | \eta \rangle
\\ \notag &+ \langle \eta | k_\nu (V^\star JU +V^\star V)  G_\nu \rangle
\\ \notag =& \langle \eta | k_\nu \eta \rangle 2\mathrm{Re}\langle \eta|G_\nu\rangle  + \mathrm{Tr}[k_\nu V^\star V] 2\mathrm{Re}\langle\eta|G_\nu\rangle + 2\mathrm{Re}\langle\eta | k_\nu U^\star JV G_\nu\rangle 
\\&+2 \mathrm{Re}\langle \eta | k_\nu V^\star V G_\nu\rangle +\langle k_\nu G_\nu | \eta \rangle.
\end{align}
Finally, for the expectation of the energy of the photon field we obtain
\begin{align}
\notag E_{\eta,B}\big( \mathrm{d}\Gamma\big[\big(\begin{smallmatrix}
|k| & 0 \\
0 & 0 
\end{smallmatrix}\big),0\big] \big)
=& \sum\nolimits_{i,j=1}^\infty \big\langle F_i \big|\big(\begin{smallmatrix}
|k| &0\\
0& 0
\end{smallmatrix}\big) F_j \big\rangle E_{\eta,B}\big(A_J(\mathcal{J}F_i)A(F_j) \big)
\\ \notag =& \sum\nolimits_{i,j=1}^\infty \big\langle F_i \big|\big(\begin{smallmatrix}
|k| &0\\
0& 0
\end{smallmatrix}\big) F_j \big\rangle 
\Big\{ 
\langle \mathcal{J} F_i | B^\star MB F_j \rangle + Q_\eta(F_i) \overline{Q_\eta(F_j)} \Big\}
\\ \notag =& 
\sum\nolimits_{i,j=1}^\infty \big\langle \varphi_i \big| \:|k| \varphi_j \big\rangle 
\Big\{ 
\langle JV \varphi_i | JV\varphi_j \rangle + \langle \varphi_j |\eta\rangle \langle \eta |\varphi_i \rangle \Big\}
\\=& \mathrm{Tr}[|k|V^\star V]+\big\langle \eta \big| |k| \eta \big\rangle.
\end{align}
By utilizing the linearity of $T\mapsto E_{\eta,B}(T)$, we can write $E_{\eta,B}\big(H_{g,\vec{0}}\big)$ as a linear combination of the above expressions
\begin{align}
\notag E_{\eta,B}\big(H_{g,\vec{0}}\big) =& \frac{1}{2} \sum\nolimits_{\nu=1}^3 \Big\{ 
\big(\mathrm{Tr}[k_\nu V^\star V ]+\langle \eta | k_\nu \eta \rangle \big)^2 
+ \mathrm{Tr}\big[(V^\star JU k_\nu)^2\big]
+ \mathrm{Tr} \big[(k_\nu V^\star V)^2\big] 
\\ \notag &+\mathrm{Tr}\big[k_\nu^2 V^\star V\big]
+ 2\langle \eta | k_\nu V^\star V k_\nu \eta \rangle
+\big\langle \eta \big| k_\nu^2 \eta \big\rangle 
+ 2\mathrm{Re}\langle \eta | k_\nu V^\star JU k_\nu \eta \rangle
\\ \notag &+ 2\langle G_\nu | V^\star V G_\nu \rangle + \|G_\nu\|^2 + 2\mathrm{Re} \langle G_\nu | V^\star JU G_\nu \rangle +4 \big(\mathrm{Re}\langle \eta | G_\nu\rangle \big)^2
\\ \notag &+\langle \eta | k_\nu \eta \rangle 2\mathrm{Re}\langle\eta|G_\nu\rangle +\mathrm{Tr}[k_\nu V^\star V] 2\mathrm{Re}\langle\eta|G_\nu\rangle + 2\mathrm{Re} \langle \eta | k_\nu U^\star JV G_\nu \rangle 
\\ \notag &+ 2\mathrm{Re} \langle \eta | k_\nu V^\star V G_\nu \rangle + \langle \eta | k_\nu G_\nu\rangle
\\ \notag &+ \langle \eta | k_\nu \eta \rangle 2\mathrm{Re}\langle \eta|G_\nu\rangle  + \mathrm{Tr}[k_\nu V^\star V] 2\mathrm{Re}\langle\eta|G_\nu\rangle + 2\mathrm{Re}\langle\eta | k_\nu U^\star JV G_\nu\rangle 
\\ \notag &+2 \mathrm{Re}\langle \eta | k_\nu V^\star V G_\nu\rangle +\langle k_\nu G_\nu | \eta \rangle \Big\} 
+\mathrm{Tr}[|k|V^\star V]+\big\langle \eta \big| |k| \eta \big\rangle
\\ =&
\notag \frac{1}{2} \sum\nolimits_{\nu=1}^3 \Big\{ 
\big(\mathrm{Tr}[k_\nu V^\star V ]+\langle \eta | k_\nu \eta \rangle +2 \mathrm{Re}\langle \eta | G_\nu \rangle \big)^2 
\\ \notag &+ \mathrm{Tr}\big[(V^\star JU k_\nu)^2\big]
+ \mathrm{Tr} [k_\nu V^\star V k_\nu (1+V^\star V)]
\\ \notag &+ \langle G_\nu +k_\nu \eta | (1+ 2V^\star V) (G_\nu +k_\nu \eta) \rangle
\\ 
&+ 2\mathrm{Re}\langle G_\nu + k_\nu\eta | V^\star JU (G_\nu +k_\nu \eta) \rangle
\Big\} +\mathrm{Tr}[|k|V^\star V]+\big\langle \eta \big|\: |k| \eta \big\rangle.
\end{align}
\end{proof}

\newpage
\section{Minimization over $V\geq 0$ and $\eta\in\mathfrak{h}$}
The goal is to show that the infimum of $E_{\eta,B}\big(H_{g,\vec{p}}\big)$ over all $U,V,\eta$ coincides with the infimum over $U,V,\eta$ of the form $U=\sqrt{1+V^2}$ with $V \geq 0$, effectively eliminating the degrees of freedom  arising from $U$. We begin by introducing a series of preperatory lemmata.

\noindent The presence of the antiunitary involution $J:\mathfrak{h}\to\mathfrak{h}$ requires us to deal with antilinear operators. Therefore we first derive an antilinear version of the Cauchy-Schwarz inequality for traces.

\Def{
Let $A:\mathfrak{h}
\to \mathfrak{h}$ be an antilinear operator with $\sup_{\|x\|=1} \|Ax\|<\infty$. The \textbf{adjoint of $A$} is the antilinear operator $A^\star:\mathfrak{h}\to\mathfrak{h}$ which fulfills \eq{
\forall x,y\in \mathfrak{h} : \la x|A y\ra = \la y|A^\star x\ra.
}
}

\Lemma{
Let $L\in \HS$ and let $A$ be an anti-linear operator on $\mathfrak{h}$ such that $A^\star A \in \mathcal{L}^1(\mathfrak{h})$. Then the Cauchy-Schwarz inequality holds true
\eq{
|\Tr(LA)| \leq \Tr(A^\star A)^\frac{1}{2} \Tr(L^\star L)^\frac{1}{2}.
}}

\begin{proof}
Let $\{\varphi_i\}_i \subseteq \mathfrak{h}$ be any orthonormal basis. Then 
\eq{
\notag &\Big|\sum\nolimits_{i\in\mathbb{N}} \la \varphi_i |LA \varphi_i \ra \Big|
\leq 
\sum\nolimits_{i\in\mathbb{N}} |\la \varphi_i |LA \varphi_i \ra| 
\\ \notag \leq &
\sum\nolimits_{i\in\mathbb{N}} \|L^\star \varphi_i\| \|A\varphi_i\|
\leq  \Big(\sum\nolimits_{i\in\mathbb{N}} \la L^\star\varphi_i|L^\star \varphi_i\ra \Big)^\frac{1}{2} \Big(\sum\nolimits_{i\in\mathbb{N}} \la A\varphi_i|A \varphi_i\ra \Big)^\frac{1}{2}
\\ =& \Big(\sum\nolimits_{i\in\mathbb{N}} \la \varphi_i|LL^\star \varphi_i\ra \Big)^\frac{1}{2} \Big(\sum\nolimits_{i\in\mathbb{N}} \la \varphi_i|A^\star A \varphi_i\ra \Big)^\frac{1}{2}
= \Tr(L^\star L)^\frac{1}{2} \Tr(A^\star A)^\frac{1}{2},
}
where various versions of the Cauchy-Schwarz inequality where used. Note that $A^\star A$ is a positive, \textit{linear} operator so that $\Tr(A^\star A)$ has a meaning. Thus, the left hand side exists for any choice of orthonormal basis and is therefore independent of the basis, see [5, Chapter VI.6]. Hence the quantity 
\eq{
\Tr(LA) := \sum_{i\in\mathbb{N}} \la \varphi_i |LA \varphi_i \ra
}
is well-defined.
\end{proof}

\noindent Additionally, we need the following technical results regarding convergence in operator norm. 

\Lemma{\label{Stetigkeit in Funktionalkalkuel}
Let $A_\infty, A_1, A_2,\dots\in\bounded$ be positive operators with $A_n \stackrel{\|\cdot\|_{op}}{\longrightarrow} A_\infty$, $n\to \infty$. Then there exists an $R>0$ such that for any $f\in C\big([0,R];\mathbb{R}\big)$
\eq{f(A_n)\stackrel{\|\cdot\|_{op}}{\longrightarrow} f(A_\infty),
} as $n\to\infty$.
}
\begin{proof}
Since norm converging sequences are bounded, there exists an $R>0$ such that \eq{
\forall n\in \mathbb{N}
\cup \{\infty\} : \|A_n\|_{op} \leq R.}
By elementary properties of the spectrum [4, Chapter VI.3] and since the involved operators are positive, we have
\eq{
\forall n\in \mathbb{N}
\cup \{\infty\} : \sigma(A_n)\subseteq [0,R].
}
Furthermore we recall that given a bounded, self-adjoint operator $A$ the functional calculus (II.13) yields 
\eq{
\forall f\in C(\sigma(A);\mathbb{R}) :\|f(A)\|_{op}\leq \|f\|_{C(\sigma(A);\mathbb{R})}.
}
Now let $f\in C([0,R];\mathbb{R})$. Since $[0,R]$ is compact there exists a sequence of polynomials $(p_m)_{m=1}^\infty$ that converges uniformly to f, by the Stone-Weierstraß Theorem. Hence for any $\varepsilon>0$ there exists a $m\in \mathbb{N}$ such that $\|f-p_m\|_{C([0,R];\mathbb{R})}< \frac{\varepsilon}{3}$ and $n_0\in\mathbb{N}$ such that for all $n \geq n_0$ 
\eq{
\notag &\|f(A_n)-f(A_\infty)\|_{op} \\ \notag \leq &\|f(A_n)-p_m(A_n)\|_{op} + \|p_m(A_n)-p_m(A_\infty)\|_{op} +\|p_m(A_\infty)-f(A_\infty) \|_{op}
\\ \leq & 2\|f-p_m\|_{C([0,R];\mathbb{R})} + \|p_m(A_n)-p_m(A_\infty)\|_{op} \leq \varepsilon, 
} since $A\mapsto p_m(A)$ is continuous.
\end{proof}

\Lemma{\label{Technical Lemma 2}
Let $B,A_\infty,A_1,A_2,\dotsc\in\bounded$ with $A_n \stackrel{\|\cdot\|_{op}}{\longrightarrow} A_\infty$, $n\to \infty$. Then \eq{A_n B A_n  \stackrel{\|\cdot\|_{op}}{\longrightarrow} A_\infty B A_\infty,
\:n\to \infty.}}
\begin{proof}
Since converging sequences are norm bounded we can see that
\eq{
\notag \|A_n B A_n-A_\infty BA_\infty\| &\leq \|(A_n-A_\infty) B A_n\| +\|A_\infty B(A_n-A_\infty)\|
\\ &\leq \|B\|(\|A_\infty\|+\sup\nolimits_{n\in\mathbb{N}} \|A_n\|) \|A_n-A_\infty\| \to 0,
} as $n\to\infty$.
\end{proof}

\subsection{Elimination of $U$}
\noindent In \cite{BH} it is shown that given $B=\big(\begin{smallmatrix} U & JVJ \\ V& JUJ \end{smallmatrix}\big)\in\Bog$  defining a Bogolubov map, we have for any $r>0$
\eq{
\notag (r+VV^\star)^{\pm 1} JU &= JU (r+V^\star V)^{\pm 1},  &(r+VV^\star)^{\pm 1} V = V (r+V^\star V)^{\pm 1},
\\(r+VV^\star)^{\pm \frac{1}{2}} JU &= JU (r+V^\star V)^{\pm \frac{1}{2}},  &(r+VV^\star)^{\pm \frac{1}{2}} V = V (r+V^\star V)^{\pm \frac{1}{2}}. \label{Properties from BH}
}
\noindent Moreover, note that given an antiunitary involution $J:\mathfrak{h}\to\mathfrak{h}$ the identity $(JAJ)^\star=JA^\star J$ holds true for any $A \in \bounded$.
The technique used in the proofs of the following two estimates is the same as in the proof of [4, Theorem IV.8].

\Lemma{ Let $B= \big(\begin{smallmatrix}
U & JVJ\\
V & JUJ
\end{smallmatrix}\big) \in \Bog$ and $\eta \in \mathfrak{h}$. Then
\eq{\big| \Real \big\la G_\nu +k_\nu \eta \big| V^\star JU (G_\nu +k_\nu \eta) \big\ra \big| \leq \big\la G_\nu +k_\nu \eta \big| \: |V| \sqrt{1+|V|^2} (G_\nu +k_\nu) \eta \big\ra.}
}
\begin{proof}
Let $
P := \| G_\nu +k_\nu \eta \|^{-2} |G_\nu +k_\nu \eta \ra \la G_\nu +k_\nu \eta |
$ and let $0\leq x \in \bounded$ be invertible.
An application of the Cauchy-Schwarz inequality yields \eq{
\notag &\| G_\nu +k_\nu \eta \|^{-2} |\Real \la G_\nu +k_\nu \eta | V^\star JU (G_\nu +k_\nu \eta )\ra | \\ \leq &  \big|\Tr \big[PV^\star  x^\frac{1}{2} x^{-\frac{1}{2}} JUP\big]\big|
\leq\Tr[PV^\star x V]^\frac{1}{2} \Tr \big[PU^\star Jx^{-1}JU\big]^\frac{1}{2}.
}
Now let $\varepsilon>0$ and \eq{x_\varepsilon := (\varepsilon +VV^\star)^{-\frac{1}{2}}(1+VV^\star)^{\frac{1}{2}}= (1+VV^\star)^{\frac{1}{2}}(\varepsilon +VV^\star)^{-\frac{1}{2}}\in\bounded .} Due to the spectral theorem the inverse of $x_\varepsilon$ is simply $x_\varepsilon^{-1} = (\varepsilon +VV^\star)^{\frac{1}{2}}(1+VV^\star)^{-\frac{1}{2}}$.
The properties $\eqref{Properties from BH}$ now lead to 
\eq{
\notag \Tr[PV^\star x_\varepsilon V] =& \Tr\big[PV^\star V(\varepsilon+V^\star V)^{-\frac{1}{2}}(1+V^\star V)^\frac{1}{2}\big]
\\ \leq & \Tr\big[P(V^\star V)^\frac{1}{2}(1+V^\star V)^\frac{1}{2}\big]
} 
as well as 
\eq{
\notag &\Tr\big[PU^\star Jx_\varepsilon^{-1}JU\big] = \Tr\big[PU^\star U (1+V^\star V)^{-\frac{1}{2}}(\varepsilon+V^\star V)^\frac{1}{2}\big]
\\ =& \Tr\big[P(1+V^\star V)^{\frac{1}{2}}(\varepsilon+V^\star V)^\frac{1}{2}\big]\rightarrow \Tr\big[P(1+V^\star V)^{\frac{1}{2}}(V^\star V)^\frac{1}{2}\big],
}
for $\varepsilon \rightarrow 0$. Note that the traces exist because $P$ is trace class. Therefore
\eq{
\notag &\| G_\nu +k_\nu \eta \|^{-2} |\Real \la G_\nu +k_\nu \eta | V^\star JU (G_\nu +k_\nu \eta )\ra | 
\\ \leq & \Tr\big[P(V^\star V)^\frac{1}{2}(1+V^\star V)^\frac{1}{2}\big]^\frac{1}{2} \Tr\big[P(1+V^\star V)^{\frac{1}{2}}(\varepsilon+V^\star V)^\frac{1}{2}\big]^\frac{1}{2}
}
for any $\varepsilon>0$ and thus when taking the limit $\varepsilon \to 0$, giving the desired inequality.
\end{proof}

\Lemma{
Let $B= \big(\begin{smallmatrix}
U & JVJ\\
V & JUJ
\end{smallmatrix}\big) \in \Bog$ and $\eta \in \mathfrak{h}$. Then
\eq{
\big|\Tr\big[(k_\nu U^\star JV)^2\big]\big| \leq \Tr\big[(V^\star V)^\frac{1}{2}(1+V^\star V )^\frac{1}{2} k_\nu (V^\star V)^\frac{1}{2}(1+V^\star V )^\frac{1}{2} k_\nu\big].
}
}
\begin{proof}
Let $\{\varphi_i\}_{i\in\mathbb{N}}$ be an orthonormal basis of $\mathfrak{h}$, $I\in \mathbb{N}$ and $\varepsilon>0$. Using $x_\varepsilon=(\varepsilon+VV^\star)^ {-\frac{1}{2}}(1+VV^\star)^{\frac{1}{2}}$ as in (IV.15) we obtain
\eq{
\notag &\Big| \sum\nolimits_{i=1}^{I} \big\la \varphi_i \big| x_\varepsilon^{-\frac{1}{2}} JU k_\nu U^\star J x_\varepsilon^{-\frac{1}{2}} x_\varepsilon^{\frac{1}{2}} Vk_\nu V^\star x_\varepsilon^{\frac{1}{2}}  \varphi_i \big\ra\Big| \\ \notag \leq & \sum\nolimits_{i=1}^{I} \big\|x_\varepsilon^{-\frac{1}{2}} JU k_\nu U^\star J x_\varepsilon^{-\frac{1}{2}} \varphi_i\big\| \big\|x_\varepsilon^{\frac{1}{2}} Vk_\nu V^\star x_\varepsilon^{\frac{1}{2}}  \varphi_i\big\| 
\\ \leq &  \Big(\sum\nolimits_{i=1}^{I} \big\| x_\varepsilon^{-\frac{1}{2}} JU k_\nu U^\star J x_\varepsilon^{-\frac{1}{2}} \varphi_i\big\|^2 \Big)^\frac{1}{2}
\Big(\sum\nolimits_{i=1}^{I} \big\la \varphi_i \big| x_\varepsilon^{\frac{1}{2}} Vk_\nu V^\star x_\varepsilon Vk_\nu V^\star x_\varepsilon^{\frac{1}{2}} \varphi_i \big\ra \Big)^\frac{1}{2}.
}
Next, note that
\eq{
\sum\nolimits_{i=1}^{I} \big\la \varphi_i \big| x_\varepsilon^{\frac{1}{2}} Vk_\nu V^\star x_\varepsilon Vk_\nu V^\star x_\varepsilon^{\frac{1}{2}} \varphi_i \big\ra \leq \Tr\big[x_\varepsilon^{\frac{1}{2}} Vk_\nu V^\star x_\varepsilon Vk_\nu V^\star x_\varepsilon^{\frac{1}{2}}\big]
}
and that as a quadratic form 
\eq{
V^\star x_\varepsilon V = \frac{V^\star V}{(\varepsilon+V^\star V)^\frac{1}{2}}(1+V^\star V)^\frac{1}{2} \leq (V^\star V)^\frac{1}{2} (1+V^\star V)^\frac{1}{2}=:A
}
due to the spectral theorem. Hence 
\eq{
\notag &\sum\nolimits_{i=1}^{I} \big\la \varphi_i \big| x_\varepsilon^{\frac{1}{2}} Vk_\nu V^\star x_\varepsilon Vk_\nu V^\star x_\varepsilon^{\frac{1}{2}} \varphi_i \big\ra \leq \sum\nolimits_{i=1}^\infty \big\la k_\nu V^\star x_\varepsilon^{\frac{1}{2}} \varphi_i \big| V^\star x_\varepsilon V (k_\nu V^\star x_\varepsilon^{\frac{1}{2}} \varphi_i) \big\ra
\\ \leq & \Tr\big[x_\varepsilon^{\frac{1}{2}} V k_\nu A k_\nu V^\star x_\varepsilon^{\frac{1}{2}}\big]= \Tr \big[A^\frac{1}{2} k_\nu V^\star x_\varepsilon V k_\nu A^\frac{1}{2}\big] \leq \Tr [A k_\nu A k_\nu].}
The above traces of positive operators exist which can be seen by using the Cauchy-Schwarz inequality
\eq{
\Tr[Ak_\nu Ak_\nu] \leq \Tr\big[k_\nu (1+V^\star  V)^\frac{1}{2} V^\star V (1+V^\star V)^\frac{1}{2} k_\nu\big] < \infty,
}
since $V^\star V \in \mathcal{L}^1(\mathfrak{h})$ and since $\mathcal{L}^1(\mathfrak{h})\subseteq \bounded$ is an ideal.
Moreover by Lemma IV.2 
\eq{
(\varepsilon+V^\star V)^\frac{1}{4} \stackrel{\|\cdot\|_{op}}{\longrightarrow} V^\star V,
} 
as $\varepsilon \to 0$, and therefore $x_\varepsilon^{-\frac{1}{2}} \stackrel{\|\cdot\|_{op}}{\to} x_0^{-\frac{1}{2}}$, as $\varepsilon \longrightarrow 0$. Furthermore by Lemma IV.3 
\eq{
x_\varepsilon^{-\frac{1}{2}} JU k_\nu U^\star J x_\varepsilon^{-\frac{1}{2}} \stackrel{\|\cdot\|_{op}}{\longrightarrow} x_0^{-\frac{1}{2}} JU k_\nu U^\star J x_0^{-\frac{1}{2}}, 
}
which, inserted in equation (IV.20) together with (IV.23) results in
\eq{
\notag &\Big| \sum\nolimits_{i=1}^{I} \big\la \varphi_i \big| x_\varepsilon^{-\frac{1}{2}} JU k_\nu U^\star J x_\varepsilon^{-\frac{1}{2}} x_\varepsilon^{\frac{1}{2}} Vk_\nu V^\star x_\varepsilon^{\frac{1}{2}}  \varphi_i \big\ra\Big|
\\ \leq & \Big( \Tr[ A k_\nu A k_\nu] \Big)^\frac{1}{2} \Big(\sum\nolimits_{i=1}^{I} \big\|x_0^{-\frac{1}{2}} JU k_\nu U^\star J x_0^{-\frac{1}{2}} \varphi_i\big\|^2 \Big)^\frac{1}{2}
}
when taking the limit $\varepsilon \to 0$ for any $I\in \mathbb{N}$. Once again by employing $\eqref{Properties from BH}$ we obtain
\eq{
U^\star J x_0^{-1} JU = (V^\star V)^\frac{1}{2} (1+V^\star V )^\frac{1}{2} = A, 
}
so that 
\eq{
\notag &\sum\nolimits_{i=1}^{I} \big\|x_0^{-\frac{1}{2}} JU k_\nu U^\star J x_0^{-\frac{1}{2}} \varphi_i\big\|^2 \leq \Tr\big[x_0^{-\frac{1}{2}} JU k_\nu U^\star J x_0^{-1} JU k_\nu U^\star J x_0^{-\frac{1}{2}}\big]
\\ =& \Tr\big[(U^\star J x_0^{-1} JU) k_\nu (U^\star J x_0^{-1} JU) k_\nu\big] = \Tr[A k_\nu A k_\nu] < \infty.
}
This implies
\eq{
\Big| \sum\nolimits_{i=1}^{I} \big\la \varphi_i \big| x_\varepsilon^{-\frac{1}{2}} JU k_\nu U^\star J x_\varepsilon^{-\frac{1}{2}} x_\varepsilon^{\frac{1}{2}} Vk_\nu V^\star x_\varepsilon^{\frac{1}{2}}  \varphi_i \big\ra\Big| \leq \Tr[A k_\nu A k_\nu],
}
for any $I\in \mathbb{N}$ and consequently when taking the limit $I\to \infty$
\eq{
\big|\Tr\big[(k_\nu U^\star V)^2\big]\big| = \Tr \big[x_\varepsilon^{-\frac{1}{2}} JU k_\nu U^\star J x_\varepsilon^{-\frac{1}{2}} x_\varepsilon^{\frac{1}{2}} Vk_\nu V^\star x_\varepsilon^{\frac{1}{2}}\big] \leq \Tr[A k_\nu A k_\nu ].
}
\end{proof}

\subsection{Estimate on the Bogolubov-Hartree-Fock-Energy}
After establishing the inequalities (IV.13) and (IV.19) the central result of section IV may be proven easily. It is of interest however that up until this point the antiunitary involution $J$ has been arbitrary. The authors of \cite{BH} place great importance on the choice \eq{J:\mathfrak{h}\to \mathfrak{h}, \psi(k)\mapsto \overline{\psi(-k)}} and make use of its properties from the start. Here, the choice of the antiunitary involution $J$ is only used for showing the second inequality in (IV.33).

\Theorem{\label{Estimate Groundstate Energy}
Let $H_{g, \vec{0}}$ be the fiber of total momentum $\vec{p}=\vec{0}$ of the Pauli-Fierz Hamiltonian $\eqref{Pauli-Fierz Hamiltonian}$. Furthermore let $J:\mathfrak{h}\to \mathfrak{h}$ be the antiunitary map $\psi(k) \mapsto \overline{\psi(-k)}$. Then
\begin{align}
\notag 
& \inf\big\{\tilde{\mathcal{E}}_{g,\vec{0}}(V, \eta) | V \in \mathcal{L}^2(\mathfrak{h}), V\geq 0,\eta \in \mathfrak{h}\big\}
\\ \notag \leq &\inf\big\{E_{\eta,B}(H_{g,\vec{0}}) | B = \big(\begin{smallmatrix}
U & JVJ
\\V & JUJ
\end{smallmatrix}\big) \in \mathrm{Bog}_J[\mathfrak{h}], \eta \in \mathfrak{h}\big\}
\\ \leq& \inf\big\{\tilde{\mathcal{E}}_{g,\vec{0}}(V, \eta) | V=JVJ \in \mathcal{L}^2(\mathfrak{h}), V\geq 0 , \eta = J\eta \in \mathfrak{h}\big\}, 
\end{align}
where $E_{\eta,B}(H_{\alpha,\vec{0}})=\langle \Omega| \mathbb{U}_B \mathbb{W}_\eta H_{\alpha,\vec{0}} \mathbb{W}_\eta^\star \mathbb{U}_B^\star \Omega \rangle$ and 
\begin{align}
\notag \tilde{\mathcal{E}}_{g,\vec{0}}(V,\eta) =& 
\frac{1}{2} \sum\nolimits_{\nu=1}^3 \Big\{ 
\big(\mathrm{Tr}\big[k_\nu V^2 \big]+\langle \eta | k_\nu \eta \rangle +2 \mathrm{Re}\langle \eta | G_\nu \rangle \big)^2 
\\ \notag &- \mathrm{Tr}\big[(k_\nu V\sqrt{1+V^2} )^2\big]
+ \mathrm{Tr} \big[k_\nu V^2 k_\nu (1+V^2)\big]
\\ \notag &+ \big\langle G_\nu +k_\nu \eta \big| \big(V-\sqrt{1+V^2}\big)^2 (G_\nu +k_\nu \eta) \big\rangle
\Big\} \\& +\mathrm{Tr}\big[|k|V^2\big]+\big\langle \eta \big| |k| \eta \big\rangle.
\end{align}
}

\begin{proof}
The first inequality in (IV.33) is shown by employing Lemma IV.5 and Lemma IV.6 in order to obtain
\eq{
\notag E_{\eta,B}\big(H_{g,\vec{0}}\big) =&
\frac{1}{2} \sum\nolimits_{\nu=1}^3 \Big\{ 
\big(\mathrm{Tr}[k_\nu V^\star V ]+\langle \eta | k_\nu \eta \rangle +2 \mathrm{Re}\langle \eta | G_\nu \rangle \big)^2 
\\ \notag &+ \mathrm{Tr}\big[(V^\star JU k_\nu)^2\big]
+ \mathrm{Tr} [k_\nu V^\star V k_\nu (1+V^\star V)]
\\ \notag &+ \langle G_\nu +k_\nu \eta | (1+ 2V^\star V) (G_\nu +k_\nu \eta) \rangle
\\ 
\notag &+ 2\mathrm{Re}\langle G_\nu + k_\nu\eta | V^\star JU (G_\nu +k_\nu \eta) \rangle
\Big\} +\mathrm{Tr}[|k|V^\star V]+\big\langle \eta \big| |k| \eta \big\rangle
\\ 
\notag \geq &
\frac{1}{2} \sum\nolimits_{\nu=1}^3 \Big\{ 
\big(\mathrm{Tr}\big[k_\nu |V|^2 \big]+\langle \eta | k_\nu \eta \rangle +2 \mathrm{Re}\langle \eta | G_\nu \rangle \big)^2 
\\ \notag &- \Tr\big[ \big(|V|\sqrt{1+|V|^2} k_\nu \big)^2 \big]
+ \mathrm{Tr} \big[k_\nu |V|^2 k_\nu \big(1+|V|^2\big)\big]
\\ \notag  &+ \big\langle G_\nu +k_\nu \eta \big| \big(1+ 2|V|^2\big) (G_\nu +k_\nu \eta) \big\rangle
\\ \notag &-2 \big\la G_\nu +k_\nu \eta \big| \: |V| \sqrt{1+|V|^2} (G_\nu +k_\nu) \eta \big\ra
\Big\} \\ &+\mathrm{Tr}\big[|k|\: |V|^2\big]+\big\langle \eta \big| |k| \eta \big\rangle
 = \tilde{\mathcal{E}}_{g,\vec{0}}(|V|,\eta).
}
Hence we have
\eq{
\notag &\inf\big\{\tilde{\mathcal{E}}_{g,\vec{0}}(V, \eta) | V \in \mathcal{L}^2(\mathfrak{h}), V\geq 0,\eta \in \mathfrak{h}\big\}
\\ \leq &\inf\big\{E_{\eta,B}(H_{g,\vec{0}}) | B = \big(\begin{smallmatrix}
U & JVJ
\\V & JUJ
\end{smallmatrix}\big) \in \mathrm{Bog}_J[\mathfrak{h}], \eta \in \mathfrak{h}\big\}.
}
For the second inequality assume $V\in \HS, \eta \in \mathfrak{h}$ with $V=JVJ=|V|,\eta=J\eta$ and set $U=\sqrt{1+V^2}$. Since $V$ commutes with $J$ it is easy to see that for any $r>0$
\eq{
(r+V)^{-1}J=(r+V)^{-1}J(r+V)(r+V)^{-1}=J(r+V)^{-1}.
}
Furthermore, the operator identity $A^{-\frac{1}{2}} = \frac{1}{\pi} \int_0^\infty (s+A)^{-1} \frac{\mathrm{d}s}{s^{1/2}}$ [4, p.26] implies 
\eq{
(r+V)^{\pm\frac{1}{2}}J=J(r+V)^{\pm\frac{1}{2}},
}
for any $r>0$. Using (IV.38) we obtain
\eq{
\notag &\Tr\big[(k_\nu U^\star JV)^2\big]= \Tr\big[k_\nu \sqrt{1+V^2}JV k_\nu \sqrt{1+V^2}JV\big] \\
=& \Tr\big[k_\nu V\sqrt{1+V^2} J k_\nu J V\sqrt{1+V^2}\big]=-\Tr\big[k_\nu V\sqrt{1+V^2} k_\nu V\sqrt{1+V^2}\big],
} and
\eq{
\notag &\Real \big\la G_\nu+k_\nu \eta \big|U^\star JV(G_\nu+k_\nu\eta) \big\ra = \Real \big\la G_\nu+k_\nu \eta \big|\sqrt{1+V^2} JV(G_\nu+k_\nu\eta) \big\ra
\\ \notag  =& \Real \big\la G_\nu+k_\nu \eta \big|V\sqrt{1+V^2}(JG_\nu +Jk_\nu\eta) \big\ra \\=&-\big\la G_\nu+k_\nu \eta \big|V\sqrt{1+V^2}(G_\nu+k_\nu\eta) \big\ra,
} where we take advantage of our choice of $J$ by virtue of $Jk_\nu J=-k_\nu,J\eta =\eta$ and $ JG_\nu=-G_\nu$. Similarly to (IV.35) applying (IV.39) and (IV.40) for the case $B=\Big(\begin{smallmatrix}
\sqrt{1+V^2} & V \\
V & \sqrt{1+V^2}
\end{smallmatrix}\Big)\in \Bog, V\geq 0$ and $\eta\in \mathfrak{h},J\eta=\eta$ yields
\eq{
E_{\eta, B}(H_{g,\vec{0}})=\tilde{\mathcal{E}}_{g,\vec{0}}(V,\eta),
} meaning that the right hand side of (IV.32) is just the infimum of $E_{\eta,B}(H_{g,\vec{0}})$ on a subset of $\mathfrak{h} \times \Bog$, which implies
\eq{
\notag &\inf\big\{E_{\eta,B}(H_{g,\vec{0}}) | B = \big(\begin{smallmatrix}
U & JVJ
\\V & JUJ
\end{smallmatrix}\big) \in \mathrm{Bog}_J[\mathfrak{h}], \eta \in \mathfrak{h}\big\}
\\ \leq& \inf\big\{\tilde{\mathcal{E}}_{g,\vec{0}}(V, \eta) | V=JVJ \in \mathcal{L}^2(\mathfrak{h}), V\geq 0 , \eta = J\eta \in \mathfrak{h}\big\}
}
\end{proof}
\Bem{
Ideally we would like the Bogolubov-Hartree-Fock-Energy to be equal to either the right hand side or the left hand side of (IV.33), thereby simplifying the minimized functional and the domain on which the minimization takes place. Indeed the choices of the antiunitary involution $J$ and the requirements $JVJ=V, J\eta=\eta, JG_\nu=G_\nu$ have been made only for saturating the inequalities (IV.13) and (IV.19) in (IV.39) and (IV.40). However, even without equality, the upper bound given by the right hand side of (IV.33) may still prove to be useful. Also, equality throughout (IV.33) yields heavy simplification of $\tilde{\mathcal{E}}_{g,\vec{0}}$ (see Section VII). 
}

\newpage
\section{Improved Parametrization}
In the proof of [4, Lemma IV.9] the parametrization
\begin{align}
V=\frac{1}{2}\big(y^\frac{1}{2}-y^{-\frac{1}{2}}\big)
\end{align}
is introduced. It is established [4, Lemma IV.10] that $y-1\geq 0$ is Hilbert-Schmidt if and only if $V$ is Hilbert-Schmidt. Furthermore the following properties hold true
\begin{align}
V^2&=\frac{1}{4y}(y-1)^2, 
\\ V\sqrt{1+V^2}&=\frac{1}{4y}(y+1)(y-1),
\\ \big(V-\sqrt{1+V^2}\big)^2 &= y^{-1}.
\end{align} 
In order to deal with Hilbert Schmidt operators, we make the substitution $z:=y-1$ which is a positive Hilbert-Schmidt operator due to [4, Lemma IV.10] and the above properties assume the form
\begin{align}
V^2&=\frac{1}{4(z+1)}z^2, 
\\ V\sqrt{1+V^2}&=\frac{1}{4(z+1)}(z+2)z,
\\ \big(V-\sqrt{1+V^2}\big)^2 &= (z+1)^{-1}.
\end{align} 
\noindent Note in passing that equations (V.4) and (V.7) yield expressions for $y$ and $z$ as functions of $V$. This parametrization yields a functional whose derivatives are easier to compute
\begin{align}
\notag \mathcal{E}_{g,\vec{0}}(z,\eta) =& 
\frac{1}{2} \sum\nolimits_{\nu=1}^3 \Big\{ 
\big(\tfrac{1}{4} \mathrm{Tr}\big[k_\nu (z+1)^{-1}z^2 \big]+\langle \eta | k_\nu \eta \rangle +2 \mathrm{Re}\langle \eta | G_\nu \rangle \big)^2 
\\ \notag &- \tfrac{1}{16}\mathrm{Tr}\big[(k_\nu (z+1)^{-1}(z+2)z)^2\big]
+ \tfrac{1}{4} \mathrm{Tr} \big[k_\nu (z+1)^{-1}z^2 k_\nu (1+\tfrac{1}{4}(z+1)^{-1}z^2)\big]
\\ \notag &+ \big\langle G_\nu +k_\nu \eta \big| (z+1)^{-1} (G_\nu +k_\nu \eta)\big \rangle
\Big\}  +\tfrac{1}{4}\mathrm{Tr}\big[|k|(z+1)^{-1}z^2\big]+\big\langle \eta \big| |k| \eta \big\rangle
\\=& 
\notag 
\frac{1}{2} \sum\nolimits_{\nu=1}^3 \Big\{ 
\big(\tfrac{1}{4}\mathrm{Tr}\big[k_\nu (z+1)^{-1} z^2 \big]+\langle \eta | k_\nu \eta \rangle +2 \mathrm{Re}\langle \eta | G_\nu \rangle \big)^2 
\\ \notag &
- \tfrac{1}{4}\mathrm{Tr} \big[k_\nu z k_\nu (z+1)^{-1} z \big]
+ \langle G_\nu +k_\nu \eta | (z+1)^{-1} (G_\nu +k_\nu \eta) \rangle
\Big\} \\& +\tfrac{1}{4}\mathrm{Tr}\big[(|k|+\tfrac{1}{2} |k|^2)(z+1)^{-1} z^2 \big]+\big\langle \eta \big| |k| \eta \big\rangle,
\end{align}
due to
\eq{
\notag & k_\nu\frac{z^2}{z+1}k_\nu \Big(4+\frac{z^2}{z+1}\Big)-k_\nu \frac{z^2+2z}{z+1}k_\nu \frac{z^2+2z}{z+1}
\\ \notag\stackrel{\Tr}{=}& 4 k_\nu^2 \frac{z^2}{z+1} - k_\nu \frac{z^2}{z+1}k_\nu \frac{2z}{z+1}- k_\nu \frac{2z}{z+1}k_\nu \frac{z^2}{z+1}- k_\nu \frac{2z}{z+1}k_\nu \frac{2z}{z+1} 
\\ \notag\stackrel{\Tr}{=}& 4 k_\nu^2 \frac{z^2}{z+1} - 4 k_\nu \frac{z}{z+1}k_\nu \frac{z}{z+1} - 4 k_\nu \frac{z^2}{z+1}k_\nu \frac{z}{z+1}
\\ \stackrel{\Tr}{=}& 4\Big(k_\nu^2 \frac{z^2}{z+1} - k_\nu z k_\nu \frac{z}{z+1} \Big),
}
where $A \stackrel{\Tr}{=} B$ means equality under trace $\Tr[A] = \Tr[B]$. Note that $\mathcal{E}_{g,\vec{0}}(z,\eta)=\tilde{\mathcal{E}}_{g,\vec{0}}\big(\frac{z}{2\sqrt{1+z}},\eta\big)$. We arrive at

\Theorem{
Let $H_{g, \vec{0}}$ be the fiber of total momentum $\vec{p}=\vec{0}$ of the Pauli-Fierz Hamiltonian $\eqref{Pauli-Fierz Hamiltonian}$ and let $J$ be defined by (IV.32). Then
\begin{align}
\notag 
& \inf\big\{\mathcal{E}_{g,\vec{0}}(z, \eta) | z \in \mathcal{L}^2(\mathfrak{h}), z\geq 0,\eta \in \mathfrak{h}\big\}
\\ \notag \leq &\inf\big\{E_{\eta,B}(H_{g,\vec{0}}) | B = \big(\begin{smallmatrix}
U & JVJ
\\V & JUJ
\end{smallmatrix}\big) \in \mathrm{Bog}_J[\mathfrak{h}], \eta \in \mathfrak{h}\big\}
\\ \leq& \inf\big\{\mathcal{E}_{g,\vec{0}}(z, \eta) | z=JzJ \in \mathcal{L}^2(\mathfrak{h}), z\geq 0 , \eta = J\eta \in \mathfrak{h}\big\}, 
\end{align}
where $E_{\eta,B}(H_{\alpha,\vec{0}})=\langle \Omega| \mathbb{U}_B \mathbb{W}_\eta H_{\alpha,\vec{0}} \mathbb{W}_\eta^\star \mathbb{U}_B^\star \Omega \rangle$ and 
\begin{align} \label{Effective Functional}
\notag \mathcal{E}_{g,\vec{0}}(z,\eta) =&   
\frac{1}{2} \sum_{\nu=1}^3 \Big\{ 
\big(\tfrac{1}{4}\mathrm{Tr}\big[k_\nu (z+1)^{-1} z^2 \big]+\langle \eta | k_\nu \eta \rangle +2 \mathrm{Re}\langle \eta | G_\nu \rangle \big)^2 
\\ \notag &
- \tfrac{1}{4}\mathrm{Tr} \big[k_\nu z k_\nu (z+1)^{-1} z \big]
+ \big\langle G_\nu +k_\nu \eta \big| (z+1)^{-1} (G_\nu +k_\nu \eta) \big\rangle
\Big\} \\& +\tfrac{1}{4}\mathrm{Tr}\big[(|k|+\tfrac{1}{2} |k|^2)(z+1)^{-1} z^2 \big]+\big\langle \eta \big| |k| \eta \big\rangle.
\end{align}
}

\newpage
\section{Euler-Lagrange-Equations}

In order to establish a necessary condition on the minimizer through variational calculus, a suitable notion of a derivative is to be defined. We choose a definition similar to the one from [3, Theorem VIII.8] which is essentially the Frechet derivative.  

\Def{Let $z\in \mathcal{L}^2(\mathfrak{h}),z\geq 0$ and $\eta \in\mathfrak{h}$. The \textbf{derivatives of $\mathcal{E}_{g,\vec{0}}(z,\eta)$} are the Hilbert-Schmidt operator $\partial_z \mathcal{E}_{g,\vec{0}}(z,\eta) \in \mathcal{L}^2(\mathfrak{h})$ and the vector $\partial_\eta \mathcal{E}_{g,\vec{0}}(z,\eta) \in \mathfrak{h}$ with the property \begin{align}
\notag &\mathcal{E}_{g,\vec{0}}(z+\delta z, \eta + \delta \eta) -\mathcal{E}_{g,\vec{0}}(z,\eta) \\=& \mathrm{Tr}[\partial_z \mathcal{E}_{g,\vec{0}}(z,\eta) \delta z] + 2\mathrm{Re} \langle \partial_\eta \mathcal{E}_{g,\vec{0}}(z,\eta) | \delta \eta \rangle + \mathcal{O}(\| (\delta z, \delta \eta) \|_{\mathcal{L}^2\times \mathfrak{h}}^2),
\end{align}
where $\eta \in \mathfrak{h}$ and where $\delta z = (\delta z)^\star \in \HS $ is chosen such that there exists an $\varepsilon>0$ so that $\mathrm{Ran}\:\delta z\subseteq \mathrm{Ran}\: \mathbbm{1}[z\geq \varepsilon\cdot \mathrm{id}_\mathfrak{h}]$, and $\|\delta z\|_{op}\leq\frac{1}{2}\varepsilon.$ Such $\delta z$ are called \textbf{variation of $z$}.
}
\Bem{The restrictions on $\delta z$ secure the positivity of $z+\delta z$.
}

\noindent Since operators in general do not commute it is not possible to single out powers of $\delta z$ as we usually would in real analysis. However

\Lemma{
Let $A,B,C \in \mathfrak{B}(\mathfrak{h})$ and let $\delta z \in \mathcal{L}^2(\mathfrak{h})\setminus \{0\}$. Then
\begin{align}
\|\delta z\|_{\mathcal{L}^2(\mathfrak{h})}^{-1} | \mathrm{Tr}(A \delta z B \delta z C) | \rightarrow 0, 
\end{align} 
as $\| \delta z\|_{\mathcal{L}^2(\mathfrak{h})} \rightarrow 0$.
}
\begin{proof}
By employing Theorem II.2 $(ii)$ and the property (II.6) of the trace we obtain \eq{
\notag &\Tr[A^\star C^\star (\delta z)^\star B^\star B \:\delta z \:C A ] \leq \|CAA^\star C^\star\|_{op}\:\|(\delta z)^\star B^\star B\: \delta z \|_1 \\
\leq& \|CAA^\star C^\star\|_{op}  \:\Tr[(\delta z)^\star B^\star B \delta z ]  \leq \|CAA^\star C^\star\|_{op} \: \|B^\star B \|_{op} \: \| \delta z \|_2^2.
} This implies that $B \:\delta z \:C A\in\HS$ and by using Cauchy-Schwarz we can see that
\eq{
\notag &\frac{1}{\|\delta z\|_2} |\Tr[A\:\delta z \: B \delta z\: CA] | \leq  \Tr[A^\star C^\star (\delta z)^\star B^\star B \:\delta z \:C A ]^\frac{1}{2} \\ \leq & \|CAA^\star C^\star\|_{op}^\frac{1}{2} \:\|B^\star B \|_{op}^\frac{1}{2}\: \| \delta z \|_2 \to 0,
} as $\| \delta z\|_2 \to 0$.
\end{proof}

\noindent Therefore when calculating derivatives we can summarize all expressions containing $\delta z$ twice under the symbol $\err$ as customary in real analysis. The next lemma is a special case of the Neumann series.

\Lemma{
Let $\delta z \in \mathcal{L}^2(\mathfrak{h})$ be a variation of $z \in \mathcal{L}^2(\mathfrak{h}),z\geq 0$ such that $\|\delta z\|_{op} <1$. Then
\begin{align}
(z+\delta z +1)^{-1} = (z+1)^{-1} \sum_{n\geq 0} (-1)^n [\delta z (z+1)^{-1}]^n,
\end{align}
where the right hand side converges in operator norm. Since $\|\cdot\|_{op}\leq \|\cdot\|_{\mathcal{L}^2}$ this equality holds whenever $\|\delta z\|_{\mathcal{L}^2}$ is small.
}

\noindent In Lemma VI.4 some commonly occuring Frechet derivatives are calculated in preparation of Theorem VI.5.

\Lemma{
Let $z\in \HS, z\geq 0$ and let $\delta z \in \HS$ be a variation of $z$. Then
\eq{
(z+\delta z)^2-z^2 &= z \: \delta z + \delta z\: z + \err,
\\(z+\delta z +1)^{-1}-(z+1)^{-1} &= -(z+1)^{-1} \delta z (z+1)^{-1} + \err
.}
}

\begin{proof}
The first identity is obvious. For the second identity we employ the previous lemma in order to obtain
\eq{
(z+\delta z +1)^{-1} -(z+1)^{-1}&= (z+1)^{-1} \notag \sum_{n\geq 1} (-1)^n [\delta z (z+1)^{-1}]^n
\\&= - (z+1)^{-1} \delta z (z+1)^{-1} + \err.
}
\end{proof}

\Theorem{
The derivatives of the functional $\mathcal{E}_{g,\vec{0}}(z,\eta)$ defined in $\eqref{Effective Functional}$ are
\eq{
\notag \partial_z \mathcal{E}_{g,\vec{0}}(z,\eta)=& \frac{1}{2} \sum_{\nu=1}^3 \Big\{\tfrac{1}{2}\big(\tfrac{1}{4}\mathrm{Tr}\big[k_\nu (z+1)^{-1}z^2 \big]+\langle \eta | k_\nu \eta \rangle +2 \mathrm{Re}\langle \eta | G_\nu \rangle \big) \\ \notag & \:\:\big[(z+1)^{-1} k_\nu z(z+1)^{-1}  +z (z+1)^{-1} k_\nu  \big]
\\ \notag &+\tfrac{1}{4}\big[ k_\nu z (z+1)^{-1} k_\nu   + (z+1)^{-1} k_\nu z k_\nu (z+1)^{-1} \big]
\\ \notag &-\big[ (z+1)^{-1}|G_\nu +k_\nu \eta \rangle \la G_\nu +k_\nu \eta | (z+1)^{-1}   \big] \Big\}
\\  \notag &+\tfrac{1}{4} \big[(z+1)^{-1} (|k|+\tfrac{1}{2}|k|^2) z (z+1)^{-1}  +z (z+1)^{-1} (|k|+\tfrac{1}{2} |k|^2) \big],
}
\eq{
\notag \partial_\eta\mathcal{E}_{g,\vec{0}}(z,\eta) =& 
\frac{1}{2} \sum_{\nu=1}^3 \Big\{ 2 \big(\tfrac{1}{4}\mathrm{Tr}\big[k_\nu (z+1)^{-1} z^2\big]+\langle\eta|k_\nu \eta\rangle +2\mathrm{Re}\langle \eta| G_\nu \rangle \big)
\big(G_\nu +k_\nu \eta  \big) 
\\  &+ k_\nu (z+1)^{-1}\big(G_\nu +k_\nu \eta\big) \Big\} + |k| \eta.
}
}

\begin{proof}
We obtain the derivative of $\Tr\big[k_\nu z^2(z+1)^{-1}\big]$ with respect to $z$ by using Lemma VI.2 and Lemma VI.4 
\eq{
&\Tr\big[k_\nu (z+\delta z)^2 (z+\delta z+1)^{-1}\big]-\Tr\big[k_\nu z^2 (z+1)^{-1}\big]
\\ \notag =& \Tr\Big\{ k_\nu \big[(z+\delta z)^2-z^2\big] (z+\delta z+1)^{-1} +k_\nu z^2 \big[(z+\delta z+1)^{-1}-(z+1)^{-1}\big] \Big\}
\\ \notag  =& \Tr\Big\{ k_\nu (z \:\delta z +\delta z\: z)(z+1)^{-1} - k_\nu z^2 (z+1)^{-1} \delta z (z+1)^{-1}
\Big\} +\err
\\ \notag =& \Tr \Big\{ \big[(z+1)^{-1} k_\nu z  + z (z+1)^{-1} k_\nu  - (z+1)^{-1} k_\nu z^2 (z+1)^{-1}\big] \delta z \Big\} +\err.
}

\noindent The derivative of $\Tr\big[(|k|+\tfrac{1}{2}|k|^2)(z+1)^{-1}z^2 \big]$ is analogous and only requires substitution of $k_\nu$ by $|k|+\tfrac{1}{2}|k|^2$. Similarly the derivative of $\la G_\nu+k_\nu \eta | (z+1)^{-1}(G_\nu +k_\nu \eta )\ra $ with respect to $z$ is computed
\eq{
\notag &\Tr\big[ |G_\nu +k_\nu \eta \rangle \la G_\nu +k_\nu \eta | (z+ \delta z +1)^{-1} \big]-\Tr\big[ |G_\nu +k_\nu \eta \rangle \la G_\nu +k_\nu \eta | (z +1)^{-1} \big]
\\ =& -\Tr\big[(z+1)^{-1} |G_\nu +k_\nu \eta \rangle \la G_\nu +k_\nu \eta | (z+1)^{-1} \delta z  \big] +\err,
}
and furthermore the derivative of $\mathrm{Tr} \big[k_\nu z k_\nu z (z+1)^{-1} \big]$ with respect to $z$ is
\eq{
\notag &\Tr\Big\{ k_\nu (z+\delta z) k_\nu (z+\delta z) (z+\delta z+1)^{-1} \Big\}-\Tr\Big\{ k_\nu z k_\nu z (z+1)^{-1} \Big\}
\\ \notag =& \Tr\Big\{ k_\nu \delta z k_\nu z (z+\delta z+1)^{-1} + k_\nu z k_\nu\delta z (z+\delta z+1)^{-1} 
\\ \notag &+ k_\nu z k_\nu z \big[(z+\delta z +1)^{-1} -(z+1)^{-1} \big]\Big\} +\err
\\ \notag =& \Tr\Big\{ k_\nu \delta z k_\nu z (z+1)^{-1} + k_\nu z k_\nu \delta z (z+1)^{-1} 
\\ \notag &- k_\nu z k_\nu z(z+1)^{-1} \delta z (z+1)^{-1}\Big\} +\err
\\ \notag =& \Tr\Big\{ \big[k_\nu z (z+1)^{-1} k_\nu +(z+1)^{-1} k_\nu z k_\nu \\&-(z+1)^{-1} k_\nu z k_\nu z(z+1)^{-1}\big] \delta z\Big\} +\err.
}
The equations (VI.10)-(VI.12) now imply 
\eq{
\notag &\mathcal{E}_{g,\vec{0}}(z+\delta z,\eta) - \mathcal{E}_{g,\vec{0}}(z,\eta) 
\\ \notag =& \frac{1}{2} \sum_{\nu=1}^3 \Big\{\tfrac{1}{2}\big(\tfrac{1}{4}\mathrm{Tr}\big[k_\nu (z+1)^{-1}z^2 \big]+\langle \eta | k_\nu \eta \rangle +2 \mathrm{Re}\langle \eta | G_\nu \rangle \big) 
\\ \notag & \:\: \Tr \big\{ \big[ \:(z+1)^{-1} k_\nu z  + z (z+1)^{-1} k_\nu  - (z+1)^{-1} k_\nu z^2 (z+1)^{-1}\big] \delta z \big\}
\\ \notag &+ \tfrac{1}{4}\Tr\big\{ \big[k_\nu z (z+1)^{-1} k_\nu +(z+1)^{-1} k_\nu z k_\nu -(z+1)^{-1} k_\nu z k_\nu z(z+1)^{-1}\big] \delta z\big\}
\\ \notag &-\Tr\big\{[ (z+1)^{-1}|G_\nu +k_\nu \eta \rangle \la G_\nu +k_\nu \eta | (z+1)^{-1} \delta z  \big\} \Big\}
\\ \notag &+\tfrac{1}{4}\Tr \big\{ \big[ \:(z+1)^{-1} (|k|+\tfrac{1}{2}|k|^2) z  + z (z+1)^{-1} (|k|+\tfrac{1}{2}|k|^2)  \\ \notag &- (z+1)^{-1} (|k|+\tfrac{1}{2}|k|^2) z^2 (z+1)^{-1}\big] \delta z \big\} +\err
\\
\notag =& \frac{1}{2} \sum_{\nu=1}^3 \Big\{\tfrac{1}{2}\big(\tfrac{1}{4}\mathrm{Tr}\big[k_\nu (z+1)^{-1}z^2 \big]+\langle \eta | k_\nu \eta \rangle +2 \mathrm{Re}\langle \eta | G_\nu \rangle \big) 
\\ \notag & \:\:\Tr \big\{ \big[(z+1)^{-1} k_\nu z(z+1)^{-1} z +z (z+1)^{-1} k_\nu \big]\delta z \big\}
\\ \notag &+\tfrac{1}{4}\Tr\big[ k_\nu z (z+1)^{-1} k_\nu \delta z  + (z+1)^{-1} k_\nu z k_\nu (z+1)^{-1} \delta z \big]
\\ \notag &-\Tr\big\{ (z+1)^{-1}|G_\nu +k_\nu \eta \rangle \la G_\nu +k_\nu \eta | (z+1)^{-1} \delta z  \big\} \Big\}
\\  \notag &+\tfrac{1}{4}\Tr \big\{\big[(z+1)^{-1} (|k|+\tfrac{1}{2}|k|^2) z (z+1)^{-1} +z (z+1)^{-1} (|k|+\tfrac{1}{2} |k|^2) \big] \delta z \big\} \\ &+\err.
}

\noindent Moreover, the Frechet derivative with regards to $\eta$ is
\eq{
\notag &\mathcal{E}_{g,\vec{0}}(z,\eta +\delta \eta)-\mathcal{E}_{g,\vec{0}}(z, \eta) 
\\ \notag =& \frac{1}{2} \sum_{\nu=1}^3 \Big\{2 
\big(\tfrac{1}{4}\Tr\big[k_\nu z^2(z+1)^{-1}\big]+\la\eta|k_\nu \eta \ra + 2\Real \la \eta | G_\nu \ra \big)
\\ \notag &\:\:\big(
\la \delta \eta|k_\nu \eta \ra +\la \eta |k_\nu  \delta\eta \ra +\la \delta \eta |G_\nu\ra + \la G_\nu |\delta \eta \ra
\big)
\\ \notag &+ \big\la G_\nu + k_\nu \eta \big| (z+1)^{-1} k_\nu \delta \eta \big\ra +\big\la k_\nu \delta \eta \big| (z+1)^{-1} (G_\nu +k_\nu \eta) \big\ra 
\Big\}
\\ \notag &+ \big\la \eta \big|\: |k| \delta\eta \big\ra + \big\la \delta\eta \big| \:|k| \eta\big\ra + \mathcal{O}\big(\|\delta\eta\|^2_{\mathfrak{h}}\big)
\\ 
=& \notag \frac{1}{2} \sum_{\nu=1}^3 \Big\{ 2
\big(\tfrac{1}{4}\Tr\big[k_\nu z^2(z+1)^{-1}\big]+\la\eta|k_\nu \eta \ra + 2\Real \la \eta | G_\nu \ra \big)
\\ \notag &\big( 2\Real 
\la k_\nu \eta |  \delta\eta \ra + 2\Real \la G_\nu |\delta \eta \ra
\big)
+ 2\Real \big\la k_\nu (z+1)^{-1}(G_\nu + k_\nu \eta)\big| \delta \eta \big\ra
\Big\}
\\  &+ 2\Real \big\la |k|\eta \: \big| \delta\eta\big\ra + \mathcal{O}\big(\|\delta\eta\|^2_{\mathfrak{h}}\big).
}
\end{proof}
\newpage
\section{Discussion of Results}

\subsection{Relation to Earlier Results}
When making the substitutions \eq{
V^\star V &= \frac{1}{2} (\mathrm{cosh}(2\hat{r})-1),\\
U^\star JV &= \frac{1}{2} \mathrm{sinh}(2\hat{r}).
} in (III.29) we obtain the same expression for the energy of a quasifree density matrix as in [3, (V.61)] 
\eq{
\notag \tilde{E}_{\eta,\hat{r}}( H_{g,\vec{0}} ) :=&
\frac{1}{2} \sum_{\nu=1}^3 \Big\{ 
\big(\mathrm{Tr}\big[\tfrac{1}{2} (\mathrm{cosh}(2\hat{r})-1) k_\nu\big]+\langle \eta | k_\nu \eta \rangle +2 \mathrm{Re}\langle \eta | G_\nu \rangle \big)^2 
\\ \notag &+\mathrm{Tr} \big[\tfrac{1}{2} (\mathrm{cosh}(2\hat{r})-1) k_\nu \tfrac{1}{2} (\mathrm{cosh}(2\hat{r})-1)k_\nu\big]
\\ \notag &+ \mathrm{Tr}\big[\tfrac{1}{2} \mathrm{sinh}(2\hat{r}) k_\nu \tfrac{1}{2} \mathrm{sinh}(2\hat{r}) k_\nu\big]
+ \mathrm{Tr} \big[k_\nu^2 \tfrac{1}{2} (\mathrm{cosh}(2\hat{r})-1)\big]
\\ \notag
&+ 2\mathrm{Re}\big\langle G_\nu + k_\nu\eta \big| \tfrac{1}{2} \mathrm{sinh}(2\hat{r}) (G_\nu +k_\nu \eta) \big\rangle
\\ \notag &+ \big\langle G_\nu +k_\nu \eta \big| \big(2 \tfrac{1}{2} (\mathrm{cosh}(2\hat{r})-1)+1\big) (G_\nu +k_\nu \eta) \big\rangle
\Big\} \\&+\mathrm{Tr}\big[\tfrac{1}{2} (\mathrm{cosh}(2\hat{r})-1)|k|\big]+\big\langle \eta \big| |k| \eta \big\rangle.
}
\Bem{
Since this representation of $E_{\eta,B}(H_{g,\vec{0}})$ is not of interest to us we will refrain from giving a detailed derivation of it. The important part is that (VII.3) coincides with the result [3, (V.61)]. 
}
\Bem{
The equations (VII.1) and (VII.2) result from the characterization of Bogolubov transformations through symplectomorphisms [3, Theorem IV.2, Definition III.6, Proposition III.7]. First, we calculate $U$ and $JV$ of a Bogolubov transformation defined through a symplectomorphism by employing the transformation behaviours [3, (IV.51)] and [3, (IV.52)] as well as making use of the polar decomposition for symplectomorphisms [3, Proposition IV.7]. Then, we use addition theorems for $\mathrm{sinh}$ and $\mathrm{cosh}$ to obtain the desired equations.
}

Furthermore since $V=\frac{z}{2\sqrt{z+1}}$ and $0=\gamma_{|\mathbb{U}_B^\star \Omega \ra\la \mathbb{U}_B^\star \Omega|} = V^\star V$ in the case of coherent states, we have $z=0$ in (V.11) and (VI.7), simplifying to 
\eq{
\notag \mathcal{E}_{g,\vec{0}}(0,\eta) =&   
\frac{1}{2} \sum_{\nu=1}^3 \Big\{ 
\big(\langle \eta | k_\nu \eta \rangle +2 \mathrm{Re}\langle \eta | G_\nu \rangle \big)^2 
+ \| G_\nu +k_\nu \eta \|^2
\Big\} \\& +\big\langle \eta \big| |k| \eta \big\rangle.
} and  \eq{
\notag \partial_\eta\mathcal{E}_{g,\vec{0}}(0,\eta) =& 
\frac{1}{2} \sum_{\nu=1}^3 \Big\{\big(\langle\eta|k_\nu \eta\rangle +2\mathrm{Re}\langle \eta| G_\nu \rangle \big)
\big(G_\nu +k_\nu \eta  \big) 
\\  &+ k_\nu \big(G_\nu +k_\nu \eta\big) \Big\} + |k| \eta.
}

\subsection{Symmetries of Minimizers}

In Theorem IV.6 the upper bound \eq{
E_{BHF}(H_{g,\vec{0}})\leq \inf\big\{\tilde{\mathcal{E}}_{g,\vec{0}}(V,\eta)|V=JVJ\in \HS,V\geq 0,\eta=J\eta\in\mathfrak{h}\big\}
} on the Bogolubov-Hartree-Fock energy has been established. Now we take a closer look at the choice $J\psi(k):=\overline{\psi(-k)}$ and the requirements $V=JVJ,J\eta=\eta$. As already mentioned at the end of section IV, these properties are only used in (IV.39) and (IV.40) for the purpose of saturating the inequalities (IV.13) and (IV.19). Furthermore in (II.58) and (II.63) we chose $\vec{G}(k)$ so that it interacts with $J$ beneficially. We can see that with $-k:=(-\vec{k},\tau)$ 
\eq{\forall \psi \in \mathfrak{h}: J k_\nu J \psi(k) = J k_\nu \overline{\psi(-k)}=-k_\nu \psi(k)
}
and \eq{
J k_\nu \eta (k) = -k_\nu \overline{\eta(-k)}=-k_\nu J\eta(k)=-k_\nu \eta(k) \\
J G_\nu(k) = g\: \big(\varepsilon_\tau(-\vec{k})\big)_\nu |\vec{k}|^{-\frac{1}{2}}=-g \: \big( \varepsilon_\tau (\vec{k})\big)_\nu |\vec{k}|^{-\frac{1}{2}}=-G_\nu(k)
}
which enabled the switch of sign in (IV.39) and (IV.40). 
\Bem{Assuming there exists another antiunitary involution $\tilde{J}$ with $\tilde{J}k_\nu=-k_\nu \tilde{J}$ there may again be a (real) choice of $\vec{\varepsilon}_\tau(\vec{k})$ such that $\tilde{J} \vec{G}=-\vec{G}$. In that case the requirements $\tilde{J}V\tilde{J}=V, \tilde{J}\eta =\eta$ may lead to minimization over a different subset of $\mathfrak{h}\times \HS$.}

\noindent Using the properties discussed above we can see that
\eq{
\la \eta | k_\nu \eta \ra = \la J\eta |k_\nu J \eta\ra = \la J k_\nu J\eta | \eta\ra = - \la \eta |k_\nu \eta\ra,
} and thus \eq{\forall \eta \in \mathfrak{h},J\eta =\eta:\la \eta | k_\nu \eta \ra=0.}
Similarly, for any $\eta \in \mathfrak{h}$ with $J\eta =\eta$
\eq{
2\Real \la \eta | G_\nu \ra = \la \eta | G_\nu\ra - \la JG_\nu | J\eta \ra = \la \eta | G_\nu\ra - \overline{\la G_\nu | \eta \ra} =0,
}
simplifying (IV.34) to
\begin{align}
\notag \tilde{\mathcal{E}}_{g,\vec{0}}(V,\eta) =& 
\frac{1}{2} \sum_{\nu=1}^3 \Big\{ 
\mathrm{Tr}\big[k_\nu V^2 \big]^2 
\\ \notag &- \mathrm{Tr}\big[\big(k_\nu V\sqrt{1+V^2} \big)^2\big]
+ \mathrm{Tr} \big[k_\nu V^2 k_\nu \big(1+V^2\big)\big]
\\ \notag &+ \big\langle G_\nu +k_\nu \eta \big| \big(V-\sqrt{1+V^2}\big)^2 (G_\nu +k_\nu \eta) \big\rangle
\Big\} \\& +\mathrm{Tr}\big[|k|V^2\big]+\big\langle \eta \big| |k| \eta \big\rangle,
\end{align}
and since $JVJ=V$ implies $JzJ=z$, equation (V.11) becomes 
\eq{\notag \mathcal{E}_{g,\vec{0}}(z,\eta) =&   
\frac{1}{2} \sum_{\nu=1}^3 \Big\{ 
\tfrac{1}{16}\mathrm{Tr}\big[k_\nu (z+1)^{-1} z^2 \big]^2 
\\ \notag &
- \tfrac{1}{4}\mathrm{Tr} \big[k_\nu z k_\nu (z+1)^{-1} z \big]
+ \big\langle G_\nu +k_\nu \eta \big| (z+1)^{-1} (G_\nu +k_\nu \eta) \big\rangle
\Big\} \\& +\tfrac{1}{4}\mathrm{Tr}\big[(|k|+\tfrac{1}{2} |k|^2)(z+1)^{-1} z^2 \big]+\big\langle \eta \big| |k| \eta \big\rangle. }
The derivatives (VI.9) of $\mathcal{E}_{g,\vec{0}}(z,\eta)$ assume the form 

\begin{align}
\notag \partial_z \mathcal{E}_{g,\vec{0}}(z,\eta)=& \frac{1}{2} \sum_{\nu=1}^3 \Big\{\tfrac{1}{32}\mathrm{Tr}\big[k_\nu (z+1)^{-1}z^2 \big]^2 \\ \notag & \:\:\big[(z+1)^{-1} k_\nu z(z+1)^{-1}  +z (z+1)^{-1} k_\nu  \big]
\\ \notag &+\tfrac{1}{4}\big[ k_\nu z (z+1)^{-1} k_\nu   + (z+1)^{-1} k_\nu z k_\nu (z+1)^{-1} \big]
\\ \notag &-\big[ (z+1)^{-1}|G_\nu +k_\nu \eta \rangle \la G_\nu +k_\nu \eta | (z+1)^{-1}   \big] \Big\}
\\  \notag &+\tfrac{1}{4} \big[(z+1)^{-1} (|k|+\tfrac{1}{2}|k|^2) z (z+1)^{-1}  +z (z+1)^{-1} (|k|+\tfrac{1}{2} |k|^2) \big],
\\
\partial_\eta\mathcal{E}_{g,\vec{0}}(z,\eta) =& 
\frac{1}{2} \sum_{\nu=1}^3 \Big\{ k_\nu (z+1)^{-1}\big(G_\nu +k_\nu \eta\big) \Big\} + |k| \eta.
\end{align}

\noindent In the case of coherent states $z=0$ we obtain 
\eq{\notag \mathcal{E}_{g,\vec{0}}(0,\eta) =&   
\frac{1}{2} \sum_{\nu=1}^3  
 \langle G_\nu +k_\nu \eta | (G_\nu +k_\nu \eta) \rangle
+\big\langle \eta \big| |k| \eta \big\rangle. 
\\ \notag =& \frac{1}{2} \sum_{\nu=1}^3 \Big\{ \|G_\nu\|^2 +2 \Real \la G_\nu | k_\nu \eta \ra \Big\}+\big\la \eta \big| (|k|+\tfrac{1}{2}|k|^2) \eta \ra 
\\ =& \frac{1}{2} \sum_{\nu=1}^3 \|G_\nu\|^2 +\big\la \eta \big| (|k|+\tfrac{1}{2}|k|^2) \eta \ra,
} where we used that $\vec{k}\cdot \vec{G}=\sum_{\nu=1}^3 k_\nu G_\nu =0$ and we obtain the derivative \eq{
\notag \partial_\eta\mathcal{E}_{g,\vec{0}}(0,\eta) =& 
\frac{1}{2} \sum_{\nu=1}^3 \Big\{ k_\nu \big(G_\nu +k_\nu \eta\big) \Big\} + |k| \eta
\\ =& (|k|+ \tfrac{1}{2} |k|^2) \eta.
}
Since for $\varphi \in \mathfrak{h}$ \eq{
(\Lambda+\tfrac{1}{2}\Lambda^2)\|\varphi\| \geq \|(|k|+\tfrac{1}{2}|k|^2) \varphi \| \geq (\sigma +\tfrac{1}{2} \sigma ^2) \|\varphi\|,
} the bounded operator $(|k|+\tfrac{1}{2}|k|^2)$ has the inverse $(|k|+\tfrac{1}{2}|k|^2)^{-1}\in\bounded$ and the necessary condition for stationarity becomes 
\eq{
\eta =0.
}
\Bem{
Equations (VII.16) and (VII.19) are special cases of [3, (VI.63)] and [3, (VI.64)].
}

\subsection{Outlook}

Throughout our considerations we always assumed the total momentum $\vec{p}=0$ to vanish. The equation [3, (V.60)] suggests that for arbitrary $|\vec{p}|\leq \frac{1}{3}$ the energy functional (V.11) may assume the form \eq{\notag \mathcal{E}_{g,\vec{p}}(z,\eta) =&   
\frac{1}{2} \sum_{\nu=1}^3 \Big\{ 
\big(\tfrac{1}{4}\mathrm{Tr}\big[k_\nu (z+1)^{-1} z^2 \big]+\langle \eta | k_\nu \eta \rangle +2 \mathrm{Re}\langle \eta | G_\nu \rangle -p_\nu \big)^2 
\\ \notag &
- \tfrac{1}{4}\mathrm{Tr} \big[k_\nu z k_\nu (z+1)^{-1} z \big]
+ \langle G_\nu +k_\nu \eta | (z+1)^{-1} (G_\nu +k_\nu \eta) \rangle
\Big\} \\& +\tfrac{1}{4}\mathrm{Tr}\big[(|k|+\tfrac{1}{2} |k|^2)(z+1)^{-1} z^2 \big]+\big\langle \eta \big| |k| \eta \big\rangle,
} 
yielding an upper bound similar to Theorem V.1 
\eq{
\notag  &\inf\big\{E_{\eta,B}(H_{g,\vec{p}}) | B = \big(\begin{smallmatrix}
U & JVJ
\\V & JUJ
\end{smallmatrix}\big) \in \mathrm{Bog}_J[\mathfrak{h}], \eta \in \mathfrak{h}\big\}
\\ \leq& \inf\big\{\mathcal{E}_{g,\vec{p}}(z, \eta) | z=JzJ \in \mathcal{L}^2(\mathfrak{h}), z\geq 0 , \eta = J\eta \in \mathfrak{h}\big\}.
} 
Assuming that we manage to calculate the right side of (VII.21) as a function $f_{\vec{p}}(\Lambda)$ of the ultraviolet cutoff $\Lambda$ for every $\vec{p}$, then we would end up with an upper bound to the Bogolubov-Hartree-Fock energy $E_{BHF}(H_{g,\vec{p}})$ and therefore to the ground state energy $E_{gs}(H_{g,\vec{p}})$ 
\eq{
E_{gs}(H_{g,\vec{p}})\leq E_{BHF}(H_{g,\vec{p}}) \leq f_{\vec{p}}(\Lambda),
}
for every fiber operator. By the properties of the direct integral decomposition [6, Theorem XIII.85] a real number $a\in \mathbb{R}$ lies in the spectrum $\sigma\big(\tilde{H}_g\big)$ of the Pauli-Fierz Hamiltonian if and only if for all $\varepsilon >0$,
\eq{
\mathrm{Vol}(\{\vec{p}\:|\:\sigma(H_{g,\vec{p}})\cap (a-\varepsilon, a+\varepsilon)\neq \emptyset \})>0.
}
Now, turning (VII.22) and (VII.23) into an upper bound $E_{gs}\big(\tilde{H}_g \big)\leq f(\Lambda)$ on the ground state energy of the Pauli-Fierz Hamiltonian may lead to new insight on its growth behaviour asymtotically in the ultraviolett cutoff $\Lambda$.

\newpage

\textbf{Erratum:} There is a technical error in the proof of Lemma IV.5, which, however, can be fixed.

\printbibliography


\end{document}